\documentclass[aps,prb,twocolumn,10pt,amsmath,amssymb,superscriptaddress]{revtex4-1}

\usepackage{graphicx}
\usepackage{dcolumn}
\usepackage{bm}
\usepackage{color}
\usepackage{epstopdf}
\usepackage{array}
\usepackage{braket}

\newcommand{\Li}{Li$_2($Li$_{1-x}$Fe$_x$)N\,}

\newcommand{\cel}[1]{\ensuremath{#1\,^\circ\textrm{C}}}
\begin{document}
\preprint{APS/123-QED}
\title{Magnetic field tuning of low energy spin dynamics in the single-atomic magnet \Li}
\author{S. A. Br\"auninger}
\affiliation{Institute of Solid State and Materials Physics, TU Dresden, D-01069 Dresden, Germany}
\author{A. Jesche}
\affiliation{Institute of Physics, University Augsburg, D-86135 Augsburg, Germany}
\author{S. Kamusella}
\affiliation{Institute of Solid State and Materials Physics, TU Dresden, D-01069 Dresden, Germany}
\author{F. Seewald}
\affiliation{Institute of Solid State and Materials Physics, TU Dresden, D-01069 Dresden, Germany}
\author{M. Fix}
\affiliation{Institute of Physics, University Augsburg, D-86135 Augsburg, Germany}
\author{R. Sarkar}
\affiliation{Institute of Solid State and Materials Physics, TU Dresden, D-01069 Dresden, Germany}
\author{A. A. Zvyagin}
\affiliation{Max-Planck-Institute for the Physics of Complex Systems, N\"othnitzer
Str. 38, D-01187 Dresden, Germany}
\affiliation{B.I.~Verkin Institute for Low Temperature Physics and
Engineering of the National Academy of Sciences of Ukraine,
Nauky Ave. 47, 61103 Kharkiv, Ukraine}
\affiliation{V.N. Karazin Kharkiv National University, Svobody sq. 4, 61022 Kharkiv, Ukraine}
\author{H.-H. Klauss}
\altaffiliation[]{henning.klauss@tu-dresden.de}
\affiliation{Institute of Solid State and Materials Physics, TU Dresden, D-01069 Dresden, Germany}
\date{\today}
\begin{abstract}
We present a systematic  $^{57}$Fe-M\"ossbauer study on highly diluted Fe centers in \Li single-crystals as a function of temperature and magnetic field applied transverse and longitudinal with respect to the single-ion anisotropy axis.
Below 30 K the Fe centers exhibit a giant magnetic hyperfine field of $\bar{B}_A=70.25(2)$\,T parallel to the axis of strongest electric field gradient $\bar{V}_{zz}=-154.0(1)$\,V/\AA$^2$. Fluctuations of the magnetic hyperfine field are observed between 50\,K and 300\,K and described by the Blume two-level relaxation model. From the temperature dependence of the fluctuation rate an Orbach spin-lattice relaxation process is deduced.
An Arrhenius analysis yields a single thermal activation barrier of $\bar{E}_A=570(6)$\,K and an attempt frequency $\bar{\nu}_0=309(10)$\,GHz. M\"ossbauer spectroscopy studies with applied transverse magnetic fields up to 5\,T reveal a large increase of the fluctuation rate by more than one order of magnitude.
In longitudinal magnetic fields a splitting of the fluctuation rate into two branches is observed consistent with a Zeeman induced modification of the energy levels. The experimental observations are qualitatively reproduced by a single-ion effective spin Hamiltonian analysis  assuming a Fe$^{1+}$ $d^7$ charge state with unquenched orbital moment and a $J=7/2$ ground state. It is demonstrated that  a weak axial single-ion anisotropy $D$ of the order of a few Kelvin can cause a two orders of magnitude larger energy barrier for longitudinal spin fluctuations.

\end{abstract}

\maketitle

\section{Introduction}
Single-molecule magnets (SMMs) and single-atomic magnets (SAMs) are model systems to study fundamental aspects of magnetic relaxation including quantum tunneling of magnetization and quantum coherence in nanoscale systems\cite{ANIE:ANIE200390099,gatteschi_molecular_2006}. Moreover, SMMs and SAMs are proposed for novel data storage devices in spintronics or possible applications for quantum computing~\cite{bogani_molecular_2008,strandberg_transition-metal_2007,leuenberger_quantum_2001}.

Isolated magnetic moments with strong axial anisotropy exhibit a bistability of magnetization with an energy barrier to magnetization reversal resulting in slow magnetic relaxation. Fluctuation of magnetization is driven by  spin-lattice relaxation or hyperfine coupling processes. For thermally activated fluctuations the spin relaxation time $\tau$ at a temperature $T$ is described by 
\begin{equation}
\tau =\tau_0\exp{\left(\frac{E_A}{k_BT}\right)},
\end{equation}
controlled by an effective energy barrier $E_A$, an 'attempt time' $\tau_0$, and with the Boltzmann constant $k_B$. 
The experimentally observed energy barrier $E_A$ can be very different from the magnetic anisotropy energy $E_m = D S^2$. Magnetic quantum tunneling processes via degenerate $S_z$ states above the $\pm S$ ground state can considerably reduce $E_A$. External magnetic fields applied longitudinal to the magnetic anisotropy axis can be used to tune the magnetic system through the level crossing condition for specific $S_z$ states via Zeeman interaction. However, quantum tunneling can only occur if off-diagonal elements in the spin Hamiltonian cause a mixing of these $S_z$ eigenstates and an avoided level-crossing with a finite minimum energy gap $\Delta_{m,m'}$ is generated. 
The Landau-Zener model~\cite{Zener1932} is applied to describe the magnetic quantum tunneling observed via magnetization steps in low temperature magnetization experiments, e.g. in nanomagnets such as Mn$_{12}$-ac~\cite{thomas_macroscopic_1996} or [Fe$_8$O$_2$(OH)$_{12}$(tacn)$_6$]$^{8+}$, briefly Fe$_8$~\cite{Sangregorio1997}. In an external magnetic field applied transverse to the anisotropy axis, the quantum tunneling can be modulated by quantum interference~\cite{PhysRevLett.69.3232}. 
This topological quantum phase interference, described by a Berry phase, is experimentally observed~\cite{Wernsdorfer1999}. 

\begin{figure}
\includegraphics[width=\columnwidth]{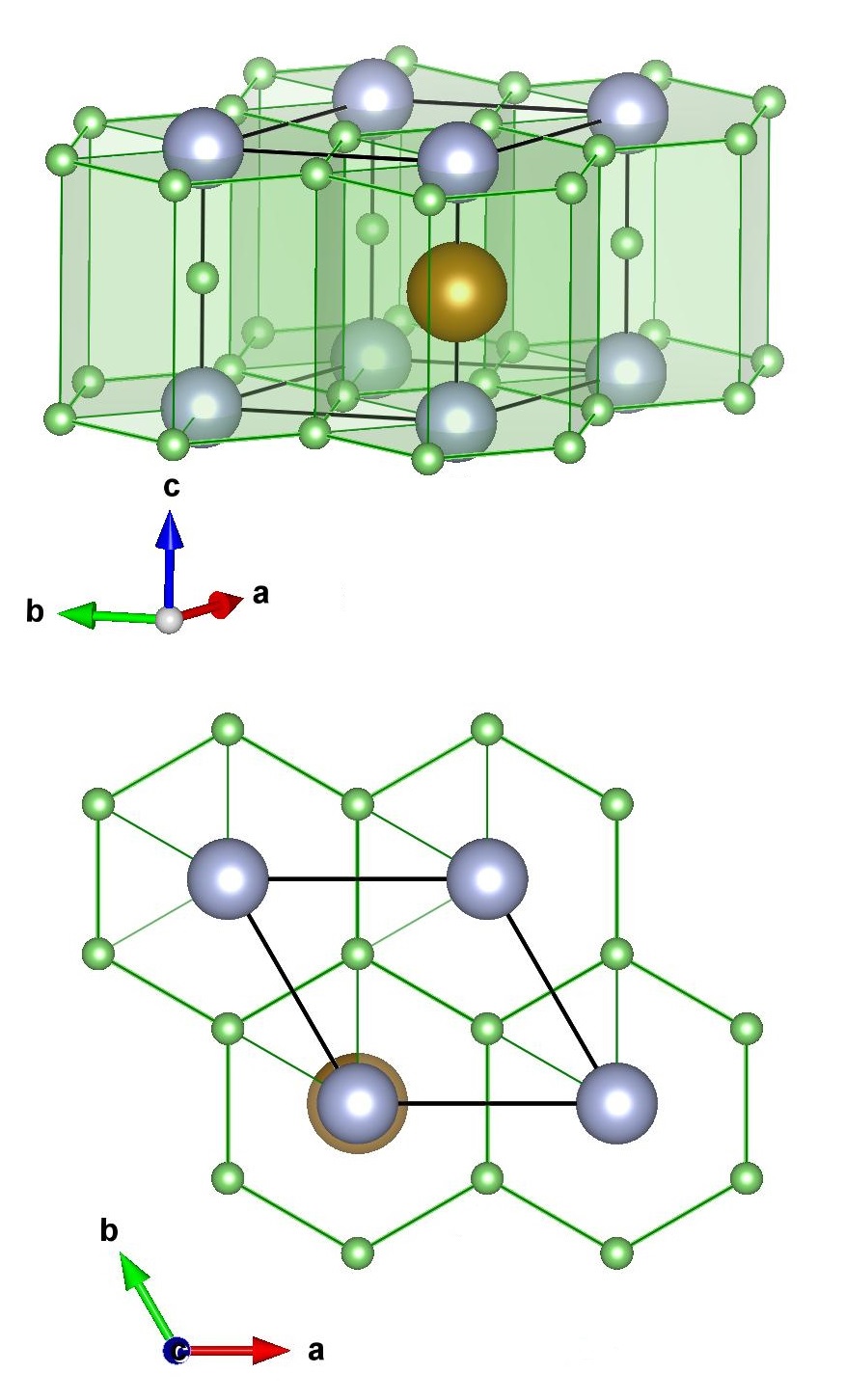}
\caption{\label{fig:cell} Crystal structure of \Li\,\cite{Klatyk1999}. The host system is an $\alpha$-Li$_3$N matrix, here Li$^+$ shown in green ($2c$ Wyckoff site, in-plane) and N$^{3-}$ in gray ($1a$ Wyckoff site).
Fe substitution, shown in brown, takes place only between the N$^{3-}$ ($1b$ Wyckoff site). The black lines show the primitive unit cell.}
\end{figure}

\Li crystallizes in a hexagonal symmetry (space group P6/mmm) and alternating planes of (Li$_2$N) and (Li$_{1-x}$Fe$_x$) are stacked along the crystallographic $c$-axis~\cite{Klatyk1999}. Figure \ref{fig:cell} shows the enhanced unit cell emphasizing the hexagonal symmetry of the Fe site and the corresponding linear N-Fe-N geometry. In Li$_3$N each N$^{3-}$ ion is surrounded by eight Li$^+$ ions. Six Li$^+$ are located in-plane in a hexagonal geometry (Li-$2c$ sites). Two Li$^+$ (Li-$1b$ sites) are located between the planes leading to a hexagonal-bipyramidal geometry. The Fe ions occupy only the Li-$1b$ site in between the Li$_2$N planes.
Studies on polycrystalline samples of concentrated \Li with $x\approx0.16$ and $x\approx0.21$ by M\"ossbauer spectroscopy revealed a static hyperfine field below $T_C\approx 65$\,K and long-range ferromagnetic ordering was proposed on the basis of magnetization studies~\cite{PhysRevLett.88.207202,ZAAC:ZAAC200300135}.
More recently, magnetization studies on large single crystals of \Li with $x = 0.001$-$0.3$ were reported~\cite{Jesche2014b, Fix2018b}.
  Large magnetic moments exceeding the spin-only value with a strong axial anisotropy parallel to the c-axis are found. 
  These magnetic moments can be associated with isolated Fe ions linearly coordinated with two Nitrogen ions in covalent N-Fe-N bonds~\cite{Fix2018c}. 
From low temperature magnetization experiments on single crystals a magnetic anisotropy field of $\mu_0H\approx88$\,T ($x\approx0.0032$) was estimated together with a large effective magnetic moment $\mu_\mathrm{eff}=6.5(4)\mu_B$ per Fe atom parallel to the $c$-axis, largely independent of the Fe concentration~\cite{Jesche2014b}. 
For $x\approx0.28$ an even larger magnetic anisotropy field of $\mu_0H\approx220$\,T was reported~\cite{Jesche2014b}.
The deduced value of $\mu_\mathrm{eff}$ is in agreement with the fully spin-orbit coupled Hunds rule value of an Fe$^{1+}$ configuration~\cite{Klatyk2002,Jesche2014b}. 
The observation of steps in magnetic hysteresis loops and relaxation phenomena with an energy barrier $E_B\approx430$\,K indicate a SMM-like behavior. The relaxation time is only weakly temperature-depended below 10\,K indicating the importance of quantum tunneling in this temperature range~\cite{Fix2018c}. However, the microscopic process of the thermally excited relaxation is not known. At low Fe doping concentrations data suggest  that the spontaneous magnetization and hysteresis is not caused by a collective magnetic ordering but rather due to the strong axial magnetic anisotropy in the linear N-Fe-N moiety \cite{Jesche2014b}. 
A recent study reports a slow paramagnetic relaxation stressing the proposed ferromagnetic nature of nondiluted \Li $(x\approx0.30)$~\cite{Fix2018b}.

Xu \textit{et al.} \cite{C7NR03041J} performed electronic structure calculations for \Li which reveal large magnetic anisotropy energies of 305\,K for an Fe$^{2+}$ $d^6$ with $J=4$ configuration and 360\,K for Fe$^{1+}$ d$^7$ with $J=7/2$ configuration. Moreover, the authors propose that an Fe$^{2+}$ state could dominate at low $x\ll 1$ whereas the Fe$^{1+}$ state should play the major role at larger $x$. However, it is not clear how such strong axial anisotropy energies around 300\,K can be reconciled with the observation of electronic level crossings in the magnetic hysteresis experiments at very low magnetic fields of $\mu_0H = 0.15$, $0.55$ and $3$\,T~\cite{Jesche2014b,Fix2018c}, i.e. energy scales $\mu B/k_B$ of several Kelvin only.

To address these questions, in this manuscript we report a detailed $^{57}$Fe-M\"ossbauer investigation on single crystals of highly diluted Fe in \Li with $x = 0.0275, 0.0109, 0.0099$, and 0.0013.
The measurements were performed at temperatures 2\,K $\leq T \leq$ 300\,K in magnetic fields 0\,T $\leq B\leq$ 5\,T applied transverse and longitudinal to the crystallographic c-axis (magnetically easy-axis). Below $T = 30$\,K the Fe centers exhibit a giant magnetic hyperfine field of $\bar{B}_A=70.25(2)$\,T parallel to the axis of strongest electric field gradient $\bar{V}_{zz}=-154.0(1)$\,V/\AA$^2$.
We demonstrate that the diluted Fe ions in \Li~indeed form isolated single-ion paramagnets consistent with an Fe$^{1+}$ $d^7$ charge state and an unquenched orbital moment, i.e. total angular momentum $J = 7/2$. 
A continuous slowing down of the spin fluctuations is observed by M\"ossbauer spectroscopy below $T = 300$\,K, which can be described by a thermally activated  Orbach process with an activation barrier of $\bar{E}_A = 570(6)$\,K. 
The fluctuation rate is very sensitive to magnetic fields of the order of a few Tesla even at elevated temperatures of $T \sim 70$\,K. A quasistatic magnetic hyperfine field is observed below 50\,K. A clustering of nearest neighbor Fe ions is ruled out by studies on samples with four different $x\leq 0.028$ proving the single atomic magnet behavior. The experimental observations are qualitatively reproduced by a single-ion spin Hamiltonian analysis. 
It is demonstrated that, for dominant magnetic quantum tunneling relaxation processes, a weak axial single-ion anisotropy $D$ of the order of a few Kelvin can cause a two orders of magnitude larger energy barrier $E_B$ for longitudinal spin fluctuations.

\section{Experimental}
Four single crystals (SCs) were investigated by $^{57}$Fe-M\"ossbauer spectroscopy in this work.
The crystals were grown out of lithium rich flux\,\cite{Jesche2014c}. 
The starting materials Li$_3$N powder (Alfa Aesar, 99.4\,\%), Li granules (Alfa Aesar, 99\,\%) and Fe granules (Alfa Aesar 99.98\,\%) were mixed in a molar ratio of Li:Fe:Li$_3$N = $6-x:x:1$ with $x = 1.5,0.5,0.5$ and 0.1 for samples SC 1, SC 2, SC 3, and SC 4, respectively.
The mixtures with a total mass of roughly 1.5\,g were packed into a three-cap Nb crucible\, \cite{Canfield2001} inside an argon-filled glovebox. 
The crucibles were sealed in $\sim 0.6$\,bar Ar via arc welding and finally sealed in a silica ampule in $\sim 0.2$\,bar Ar. 
The mixtures were heated to $T$ = \cel{900} within 5\,h, cooled to $T$ = \cel{750} over 1.5\,h, slowly cooled to $T$ = \cel{500} over 60\,h and finally decanted to separate the crystals from the excess flux. The composition was determined by inductively-coupled-plasma optical-emission-spectroscopy (ICP-OES) using a Vista-MPX. To this end the samples were dissolved in a mixture of hydrochloric acid and distilled water. Obtained Fe concentrations based on the measured Li:Fe ratio are given in Table\,\ref{tab-ICP}. Magnetization measurements were performed using a 7\,T Magnetic Property Measurement System (MPMS), manufactured by Quantum Design.
\begin{table}
\begin{tabular}{c|c|c|c}
  & $x$ [\%] & $\Delta x$ [\%] & also denoted as\\
\hline
SC 1 & 2.75 & 0.16 	&	Li$_2$(Li$_{0.9725}$Fe$_{0.0275}$)N	\\
SC 2 & 1.09 & 0.07	&	Li$_2$(Li$_{0.9891}$Fe$_{0.0109}$)N	\\
SC 3 & 0.99 & 0.06 	&	Li$_2$(Li$_{0.9901}$Fe$_{0.0099}$)N	\\
SC 4 & 0.13 & 0.01  &	Li$_2$(Li$_{0.9987}$Fe$_{0.0013}$)N \\
\hline
\end{tabular}
\caption{Fe concentration $x$ and estimated error $\Delta x$ in \Li determined by ICP-OES.}
\label{tab-ICP}
\end{table}
M\"ossbauer measurements were carried out in CryoVac and Oxford instruments helium flow cryostats in under-pressure mode or normal mode, respectively. We used a WissEl M\"ossbauer spectrometer. The detector was a proportional counter tube or Si-PIN-detector from KETEK and the source a Rh/Co source with an initial activity of 1.4 GB. The superconducting coil was powered by an Oxford instruments IPS 120-10 power supply with an applied magnetic field parallel or perpendicular to the $\gamma$-beam with an angle error of $<5^\circ$. The absorber SC 1 exceeded the thin absorber limit requiring a transmission integral fit. The analysis was done using the Moessfit analysis software~\cite{2016HyInt.237...82K}. 
All measurements were performed with the $\gamma$-beam parallel parallel to the crystallographic c axis. The single crystals were protected by paraffin wax to avoid oxidation.
\section{Results}

\subsection{Macroscopic Magnetization}
\begin{figure}
\includegraphics[width=\columnwidth]{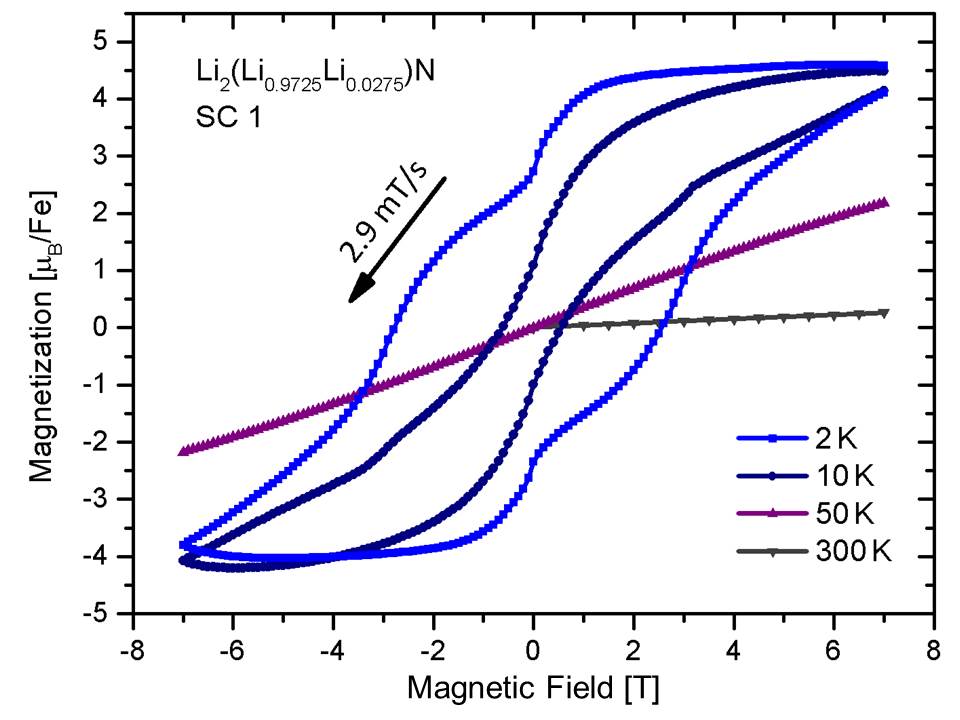}
\caption{\label{fig:m-h} Isothermal magnetization of \Li~with $x= 0.0275$ (crystal SC 1) for $B_L \parallel c$ at various temperatures given in the plot.}
\end{figure}

Figure~\ref{fig:m-h} shows the isothermal magnetization of SC 1 measured at different temperatures for magnetic field applied parallel to the crystallographic c-axis, $B_L \parallel c$. The effective sweep rate for the full loops was 2.9\,mT/s with 10\,mT/s between the measurements. Hysteresis emerges for temperatures $T < 50$\,K. At $T = 2\,\mathrm{K}$, steps appear at $\mu_0 B_L \approx 0$ and $\pm 3.3$\,T as well as for $\mu_0 B_L = 0.55$\,T, with the latter being recognizable only in the derivative d$M$/d$B$. At lower $x$, additional steps appear at $\mu_0 B_L = 0.15$\,T and the anomalies become sharper~\cite{Jesche2014b, Fix2018c}. 
The $M$-$H$ measurement shown in Fig.\,\ref{fig:m-h} was performed after the M\"ossbauer experiment and is in good agreement with results published earlier~\cite{Jesche2014b}.

\subsection{Low temperature $^{57}$Fe-M\"ossbauer spectroscopy at base temperature $T_B$}
\begin{figure}
\includegraphics[width=\columnwidth]{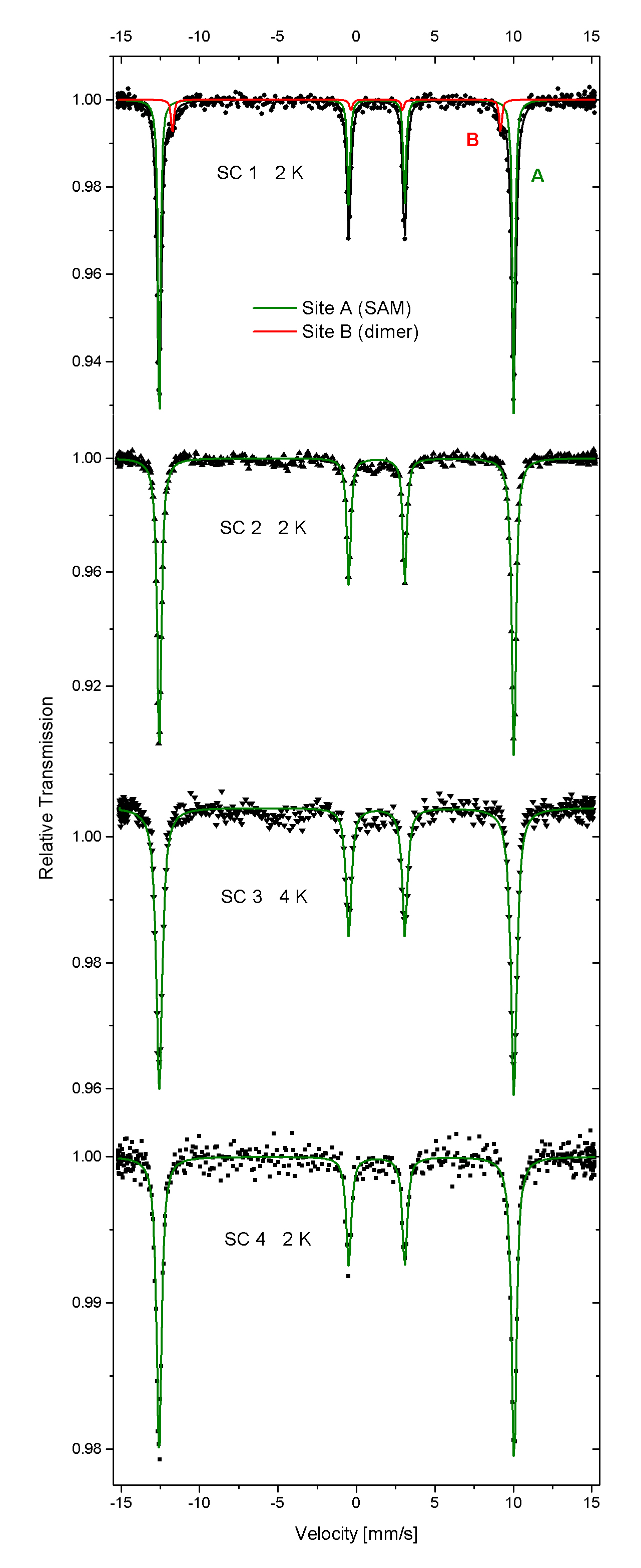}
\caption{\label{fig:Graph1}Static M\"ossbauer spectra of \Li of the single crystals SC 1-4 as specified in Table\,\ref{tab-ICP} at base temperature $T_B\leq4$\,K in ZF.}
\end{figure}
M\"ossbauer spectroscopy was performed at base temperature $T_B\leq4$\,K in zero-field (ZF) on the crystals SC 1-4. 
At this temperature the lifetime of the electronic states exceeds that of the nuclear states. Therefore, the hyperfine interactions are effectively stationary.

Fig.~\ref{fig:Graph1} shows the $^{57}$Fe-M\"ossbauer measurement at $T_B$ in ZF of SC 1-4. For SC 1, two Fe sites A (green) and B (red) are observed. 
The black line is given by the total transmission integral fitting function

\begin{equation}
T(v)=\int^{\infty}_{-\infty}L(E,v)e^{-\sigma (E)t_a}~dE
\end{equation}

with the normalized Breit-Wigner resonance cross section $\sigma(E)$ depending on the energy $E$ and an effective thickness $t_a=2.39(10)$ reflecting a non-thin absorption limit. 
Therefore, the black line represents the transmission integral fit whereas the lines for the sites A and B show the natural line $L(E,v)$.
$T(v)$ is proportional to the line intensity of the recoil-free $\gamma$-ray, a function of the absorber thickness, and $v$ is the Doppler velocity, for details see appendix. 
A thin absorber approximation is only valid for $t_a<1$ and then is the line shape described by a Lorentzian $L(E,v)$~\cite{chen_mossbauer_2007}. The fit is for SC 2-4 closer to the full thin absorption limit since the black line is replaced and shown by the green single Fe site A. 
The model used in Fig.~\ref{fig:Graph1} is the static crystal Hamiltonian $\hat{\mathcal{H}}=\hat{\mathcal{H}}_\mathrm{M}+\hat{\mathcal{H}}_\mathrm{Q}+\hat{\mathcal{H}}_\mathrm{Z}$
assuming the same electric monopole $\hat{\mathcal{H}}_\mathrm{M}$ and  quadrupole interaction $\hat{\mathcal{H}}_\mathrm{Q}$ for the two Fe sites A and B and independent Zeeman terms $\hat{\mathcal{H}}_\mathrm{Z}$. 
We deduced an isomer shift of $\delta=0.100(21)$\,mm/s with respect to $\alpha$-Fe at room temperature assuming a negligible second-order Doppler shift $\delta_\mathrm{SOD}\approx0$ of the absorber at this temperature.
The electric monopole interaction between the nuclear charge $Ze=26e$ of $^{57}$Fe with the charge number $Z$ and the $s$-electron charge density $\rho_e(0)=-e|\psi(0)|^2_{A/S}$ at the nucleus is shifted by the energy $E_A$ of the absorber material relative to the energy shift $E_S$ of the source
and leads to an effective energy shift 

\begin{equation}
\delta=\frac{2\pi}{5}Ze^2S(Z)\{|\psi(0)|^2_A-|\psi(0)|^2_S\}\left(R_{(e)}^2-R_{(g)}^2\right).
\end{equation}

Here, $R_{(g)}^2= \langle{r^2}_{(g)} \rangle$ and $R_{(e)}^2=\langle{r^2}_{(e)}\rangle$ are the mean square values of the radii of the $^{57}$Fe nucleus of the excited state (e) with nuclear spin $\frac{3}{2}$ and ground state (g) with nuclear spin $\frac{1}{2}$, respectively. $S(Z)$ is the dimensionless relativity factor which takes the spin-orbit coupling into account, e.g. for $^{57}$Fe around $S(26)\approx1.32$ or for neptunium $S(93)\approx19.4$. 
These values vary slightly depending on the oxidation state.
The monopole interaction $\hat{\mathcal{H}}_\mathrm{M}$ is given by a scalar $\hat{\mathcal{H}}_\mathrm{M}\equiv\delta+\delta_\mathrm{SOD}(T)$ as a function of the temperature. $\delta_\mathrm{SOD}$ is the second-order Doppler shift and a direct consequence of the time dilation according to the relativity theory of the lattice dynamics. The $\gamma$-photon frequency $\omega_0$ is shifted according to the transverse Doppler effect in the laboratory frame to
\begin{equation}
\omega=\omega_0\frac{\sqrt{1-v^2/c^2}}{1-v\cos\alpha/c}\approx\omega_0\left(1+\frac{v}{c}\cos\alpha-\frac{v^2}{2c^2}\right)
\end{equation}
where $v$ is the velocity of the nucleus, $\alpha$ the angle between the movement of the nucleus and $\gamma$-photon absorption and $c$ the speed of light. The last term assumes $v\ll c$. This yields in the Debye approximation the expression
\begin{equation}
\delta_\mathrm{SOD}=-C\left[{\Theta_D}+8T\left(\frac{T}{\Theta_D}\right)^3\int_0^{\Theta_D/T}\frac{x^3\mathrm{d}x}{e^x-1}\right]
\end{equation}
with
\begin{equation}
C=\frac{9k_BE_\gamma}{16M_\mathrm{eff}c^2}
\end{equation}
where $M_\mathrm{eff}$ is the effective mass. Using this expression to analyze the temperature dependence of the central shift in SC 1 yielded a Debye-temperature of $\Theta_D=315(8)$\,K. For details see appendix. 
To describe the electric quadrupole interaction $\hat{\mathcal{H}}_\mathrm{Q}$, e.g. SC 1 has a principle axis of the largest component of the EFG (electric field gradient) of $V_{zz}=-154.1(2)$\;V/\AA$^2$,
denoted as usual,
\begin{equation}
|V_{zz}|\geq|V_{xx}|\geq|V_{yy}|
\end{equation}
and the introduced asymmetry parameter
\begin{equation}
\eta=\frac{|V_{xx}|-|V_{yy}|}{|V_{zz}|}\leq 1.
\end{equation}

This leads to the reduced quadrupole Hamiltonian
\begin{equation}
\hat{\mathcal{H}}_Q=\frac{eQV_{zz}}{4I(2I-1)}[3\hat{I}^2_z-\hat{I}^2+\frac{1}{2}\eta(\hat{I}^2_++\hat{I}^2_-)]
\end{equation}
with the quadrupole moment $Q$ and the raising and lowering spin operators $\hat{I}_{\pm}=\hat{I}_x \pm i\hat{I}_y$.
 
The pure quadrupole energy eigenvalues are given by
\begin{equation}
E_Q(m)=\frac{eQV_{zz}}{4I(2I-1)}[3m^2-I(I+1)]\sqrt{1+\frac{\eta^2}{3}}
\end{equation}
with $I=\frac{3}{2}$.
The negative sign of $V_{zz}<0$ corresponds to an elongation of the EFG charge distribution and an excess of negative charges in c-axis, the elongated case of the EFG ellipsoid~\cite{gutlich_mossbauer_2011}. 
The asymmetry parameter $\eta=0$  assuming axial symmetry was used due to the hexagonal structure. 

The magnetic hyperfine or Zeeman term $\hat{\mathcal{H}}_\mathrm{Z}$ of the Hamiltonian $\hat{\mathcal{H}}_\mathrm{hf}$ is given by $\hat{\mathcal{H}}_\mathrm{Z}=-g_N\mu_N\mathbf{\hat{I}}\cdot\mathbf{\hat{B}}$
with nuclear Land\'{e} factor $g_N$, the nuclear magneton $\mu_N=e\hbar/2m_pc$, the proton mass $m_p$ and the magnetic field $\mathbf{B}$.
Taking the scalar $|\mathbf{B}|=B$ and expresses $\hat{\mathcal{H}}_\mathrm{Z}$ by the polar angle $\Theta$ and the azimutal angle $\Phi$ of $\mathbf{B}$ relative to direction of $V_{zz}$ yields
\begin{multline}
\hat{\mathcal{H}}_\mathrm{Z}=-g_N\mu_NB\bigg(\frac{\hat{I}_+e^{-i\Phi}+\hat{I}_-e^{+i\Phi}}{2}\sin\Theta\\
+\hat{I}_z\cos\Theta\bigg).
\end{multline}
The values of the magnetic hyperfine fields for the two Fe subspecies converged to $B_A=70.21(1)$\,T and $B_B=65.0(2)$\,T. Site A is the dominant Fe site. Site B  is only observed in SC 1 containing the highest Fe concentration with an  intensity fraction of 5.9(3)\%.

The two transitions
\begin{equation}
m=\pm\frac{1}{2}\rightarrow m=\pm\frac{1}{2}
\end{equation}
with $\Delta m=0$ are not observed in the spectra of Fig.\,\ref{fig:Graph1}.
The relative line intensities $W(\beta)$ depend on the polar texture angle $\beta$ describing the direction of the incident $\gamma$-ray with respect to the magnetic hyperfine field direction, $W(\beta)\propto\sin^2\beta$. For the analysis $\Theta=0$ was taken.  
The angle $\beta$ between the principle axis (largest component) $V_{zz}$ of the EFG tensor and the $\gamma$-beam was assumed to be an identical fit parameter for the monomer site A and the dimer site B. The resulting value $\beta = 0$ proves that the magnetic hyperfine field and $V_{zz}$ are aligned parallel to the c-axis.
The result is the observed ratio of the spectral line intensities of 3:0:1:1:0:3.

Fig.~\ref{fig:Graph1} shows the measurements of SC 2-4 at $T_B$. 
No indications for Fe site B are observed in SC 2 as well as in SC 3 and SC 4. 
The green line is the fit of the model of the static crystal Hamiltonian $\hat{\mathcal{H}}_\mathrm{hf}$ with an isomer shift $\delta=0.119(20)$\,mm/s and a principle axis of the EFG of $V_{zz}=-154.2(4)$\;V/\AA$^2$. The asymmetry parameter is assumed to be $\eta=0$. 
The fit yielded a magnetic hyperfine field $B_{A}=70.24(1)$\,T parallel to $V_{zz}$ of the EFG tensor and parallel to the $\gamma$-beam as well. Table\,\ref{table6} shows the obtained hyperfine parameters of SC 1-4 and the calculated mean values of $\bar{V}_{zz}$, $\bar{B}_A$, $\ln\bar{\nu}_0$ and $\bar{E}_A$. 
The hyperfine parameters are nearly concentration-independent. 
The  absolute values of the magnetic hyperfine fields $B_{A}$ and $B_{B}$ are above typical spin-only values in solid state systems and can be understood in terms of a strong unquenched orbital contribution.   
The analysis to obtain the fluctuation rate parameters $\ln\bar{\nu}_0$ and $\bar{E}_A$ of the Arrhenius temperature dependence are described in appendix \ref{App-Arrhenius}.
\begin{table}
\begin{tabular}{c|c|c|c|c}
  & $V_{zz}$ [V/\AA$^2$] & $B_A$ [T] & $E_A$ [K] &$\ln(\nu_0 [\mathrm{MHz}])$ \\
  \hline
SC 1 & -154.1(2) & 70.21(1)&552(26)&12.36(32)\\
SC 2 & -154.2(4) & 70.24(1)&563(12)&12.48(11)\\
SC 3 & -154.0(2) & 70.23(1)&581(12)&12.65(11)\\
SC 4 & -154.0(6) & 70.30(2)&552(44)&12.08(49)\\\hline
Mean value   & -154.0(1) & 70.25(2)&570(6)&12.64(7)\\
\end{tabular}
\caption{Hyperfine parameters $V_{zz}$, $B_A$ as well as the Arrhenius parameters $\ln\nu_0$ and $E_A$ of SC 1-4. The errors represent standard deviations obtained during linear regression.}
\label{table6}
\end{table}

\subsection{Zero Field $^{57}$Fe-M\"ossbauer Spectroscopy for $T_B< T <300$\,K}

\begin{figure*}
\includegraphics[width=\textwidth]{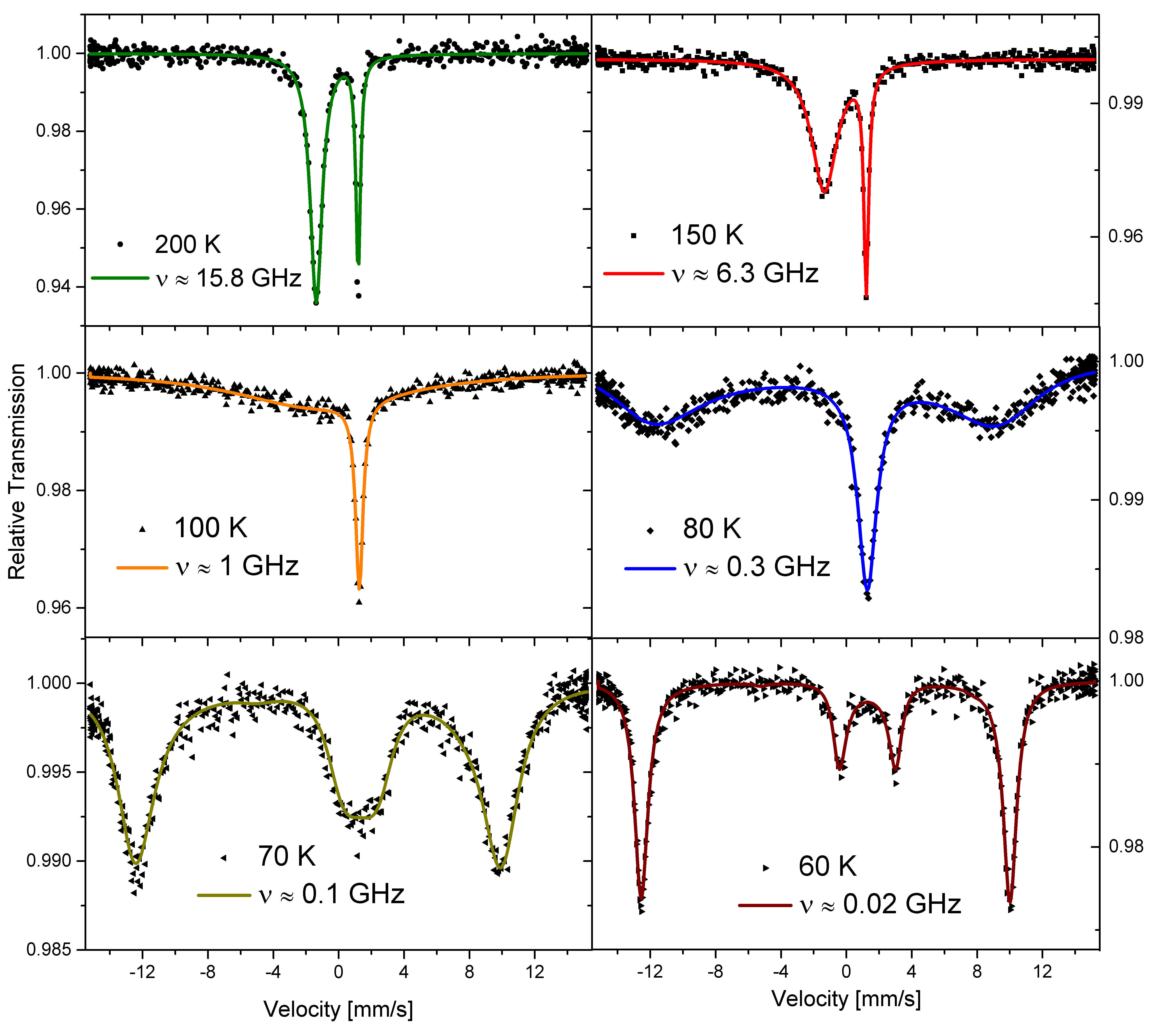}
\caption{Shown are the spectra of SC 1, \Li with $x=2.75$\%. The lines show a two level Blume model fluctuation spectrum analysis~\cite{2016HyInt.237...82K}. The magnetic hyperfine field fluctuates between $+B_A$ and $-B_A$ with the frequency $\nu$. $B_{A}$ in this model is assumed to be parallel to $V_{zz}$ and to the $\gamma$-beam.}
\label{fig:2}
\end{figure*}

Above 50\,K the $^{57}$Fe nucleus interacts with a fluctuating magnetic hyperfine field. Fig.~\ref{fig:2} shows representative spectra between 60\,K and 200\,K of SC 1 with $x=2.75$\%. In the following we will only consider Fe site A, site B is neglected in this analysis. The fit represents a Blume dynamic line shape model in the presence of quadrupole hyperfine interactions for two states, described by absorption cross section
\begin{equation}
\sigma=-\frac{\sigma_a\Gamma_0}{2}\mathrm{Im}\sum_\alpha\mathrm{Sp}\left(\hat{V}_\alpha\bra{W}\mathbf{\hat{A}}^{-1}(\omega,\Theta)\ket{1}\hat{V}_\alpha^+\right).
\end{equation}
Here, $\hat{V}_\alpha$ is the operator of hyperfine interactions of the $\gamma$-beam with polarization $\alpha$ and the nucleus, $\sigma_a$ the effective absorber thickness and $\bra{W}$ and $\ket{1}$ as described by Chuev and therein~\cite{0953-8984-23-42-426003}. The superoperator
\begin{equation}
\mathbf{\hat{A}}(\omega,\Theta)=\tilde{\omega}+\mathrm{i}\Gamma_0/2-\mathbf{\hat{L}}(\Theta)+\mathrm{i}\mathbf{\hat{P}}.
\end{equation}
is defined by the Liouville operator of hyperfine interactions $\mathbf{\hat{L}}(\Theta)$, the resonance transition energy $E_0$ is given by the corresponding frequency $\tilde{\omega}=\omega-E_0/\hbar$, $\Gamma_0$ the width of the excited nuclear level and $\mathbf{\hat{P}}$ the matrix of hyperfine transitions~\cite{PhysRevA.34.4286,2016HyInt.237...82K,0953-8984-23-42-426003,PhysRev.174.351}.\\
The initial conditions for the analysis are identical to the static case at 2\,K. A two level relaxation model was used taking into account an electronic spin reversal process. The magnetic hyperfine field fluctuates with the fluctuation frequency $\nu$ between the two values $+B_{A}$ and $-B_{A}$. 
Above 60\,K the spectral lines begin to broaden due to the fluctuations, see Fig.~\ref{fig:2}. With increasing $\nu$ at 70\,K the two internal lines collapse first  yielding a singlet at 100\,K. 
At 150\,K and above, the left resonance line of the quadrupole doublet, which is expected to appear in the fast relaxation limit $\nu \rightarrow \infty $, results from the collapse of the external lines~\cite{Carretta2006}. The Arrhenius parameter $E_A$ and $\ln\nu_0$ are obtained by an Arrhenius analysis
\begin{equation}
\ln\nu=\ln\nu_0-\frac{E_A}{k_B}\left(\frac{1}{T}\right)
\end{equation}
of the extracted fluctuation frequencies $\ln\nu$ of SC 1-4.
In this analysis, the values of $\nu$ for $T<50$\,K are not considered since these value reflect the lower bound of the fluctuation rate which the M\"ossbauer spectra analysis can resolve. 
This yielded a thermal activation barrier of $E_A=552(26)$\,K and $\ln\nu_0=12.36(32)$\,GHz for SC 1. 
The fluctuation frequency $\nu$ of Fe site A is essentially concentration-independent in SC 1-4. 
Table \ref{table6} shows the Arrhenius plot fit parameter of SC 1-4, for details see appendix.

\subsection{$^{57}$Fe-M\"ossbauer Spectroscopy in transverse magnetic Fields $B_T$} \label{MinBtrans}

In the following we present the results of systematic M\"ossbauer spectroscopy experiments under applied transverse magnet field $B_T$. These experiments were performed on sample SC 1. For eight temperatures between 30\,K and 247\,K a magnetic field up to 5\,T was applied perpendicular to the normal vector of the sample plate and therefore to the crystallographic c-axis and perpendicular to the $\gamma$-beam, see discussion. Therefore, the field was applied perpendicular to the quantization axis of the Fe spins which is identical to the low temperature orientation axis of the magnetic hyperfine field at the Fe nucleus. 
In this geometry, an increasing field $B_T$ leads to an increasing mixture of the $S_z$-eigenstates of the electronic spins and an increasing fluctuation rate of the magnetic hyperfine field is expected supported by a theoretical treatment based on the minimal spin Hamiltonian of a single-ion in the next subsection.
\begin{figure}
\includegraphics[width=\columnwidth]{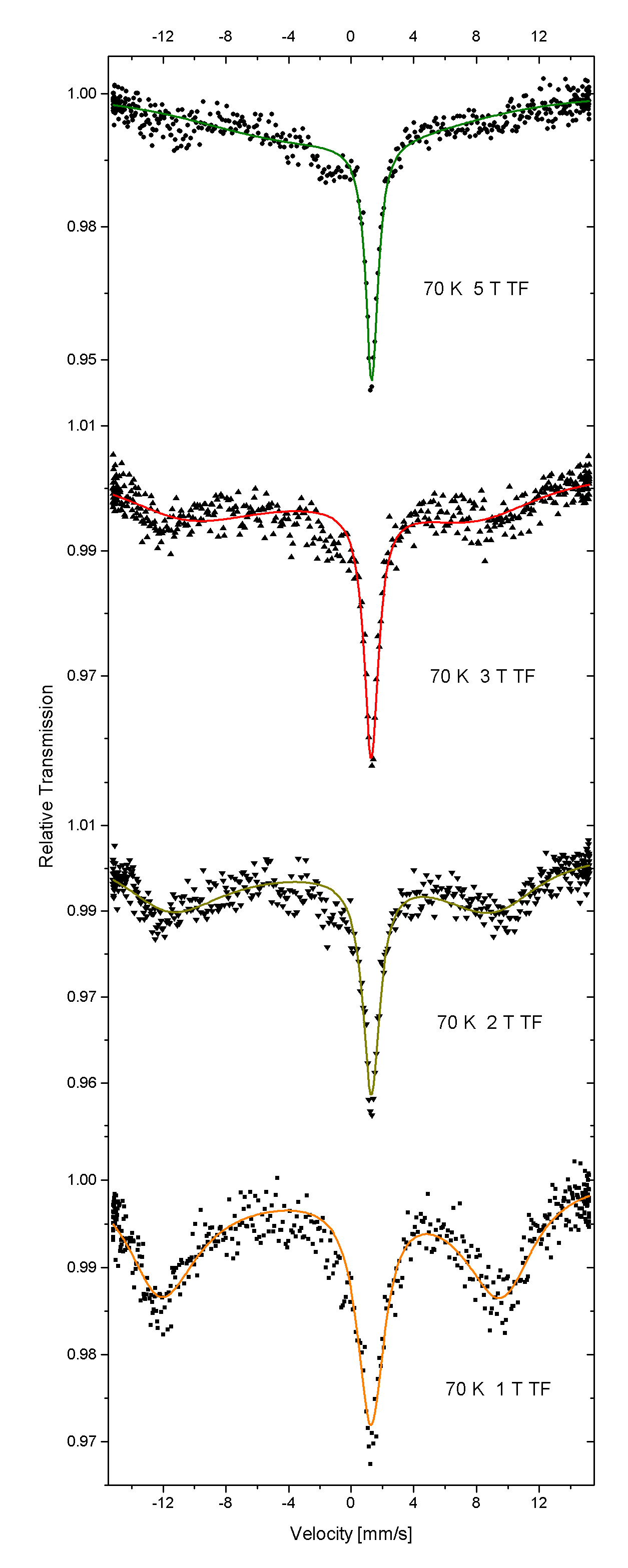}
\caption{\label{fig:field}M\"ossbauer spectra of SC 1 in various transverse magnetic fields $B_T$ at 70\,K. The corresponding ZF measurements are shown in Fig.~\ref{fig:2}.}
\end{figure}
Figure~\ref{fig:field} shows four typical M\"ossbauer spectra in different transverse magnetic fields (TF) up to 5\,T. The experimental data clearly reveal an increase of the fluctuation frequency $\nu$ with increasing field strength. The temperature and field range for these experiments was chosen such that the slowly fluctuating magnetic hyperfine field of $\approx70$\,T can be regarded as the dominant hyperfine interaction with the $^{57}$Fe nuclei and the Blume model of axial fluctuations of the magnetic hyperfine field described in the former subsection can be used for the quantitative analysis (solid lines in Fig.~\ref{fig:field}). For higher fields the vector sum of the external field and the internal magnetic hyperfine field must be considered.

\begin{figure}
\includegraphics[width=0.8\columnwidth]{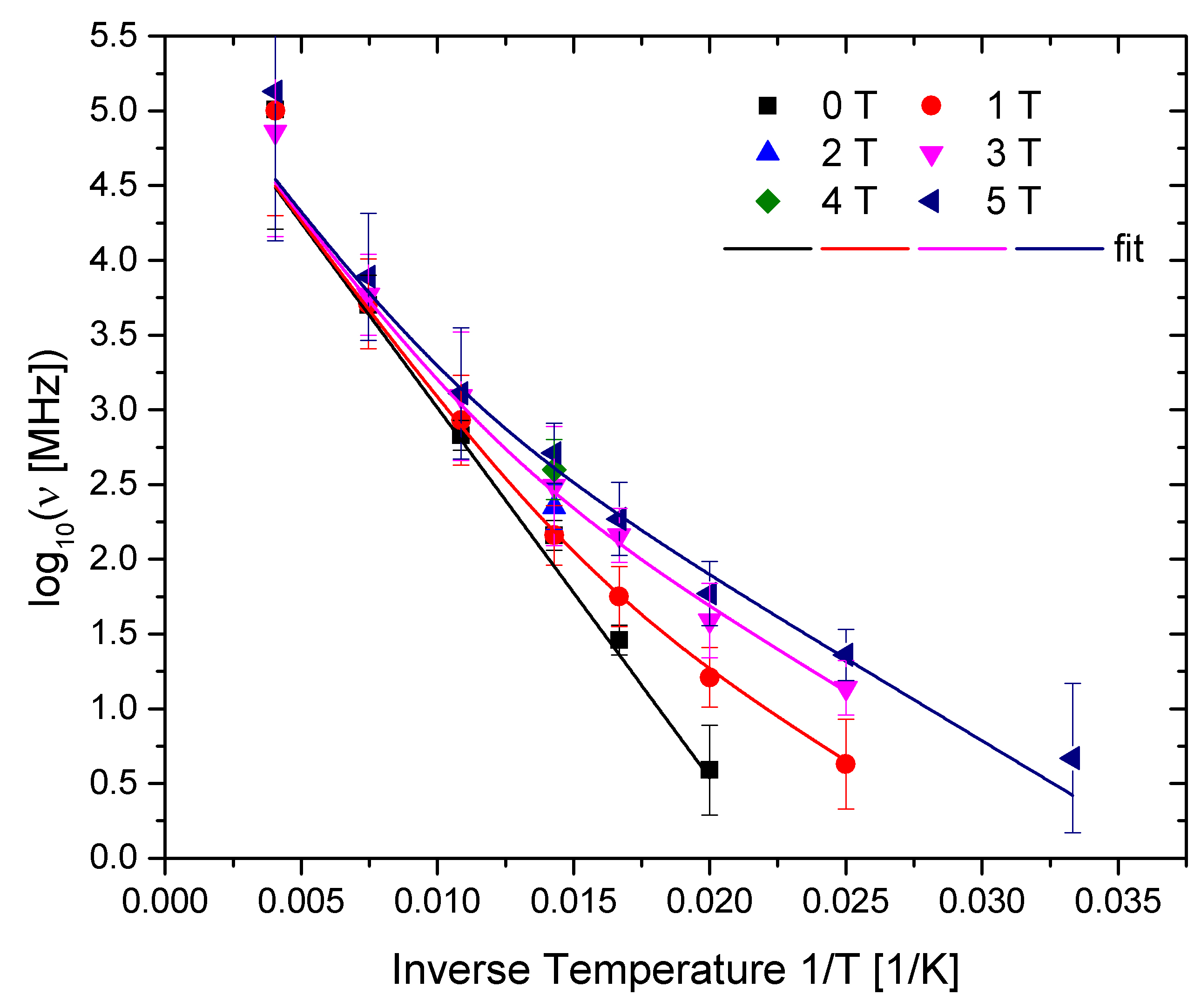}
\caption{\label{fig:Wtemperatur} Dependence of the spin fluctuation rate on temperature for \Li with $x=2.75\%$, SC 1, in transverse magnetic fields $B_T$. The solid line fitting function is described in the text.}
\end{figure}

\begin{figure}
\includegraphics[width=0.8\columnwidth]{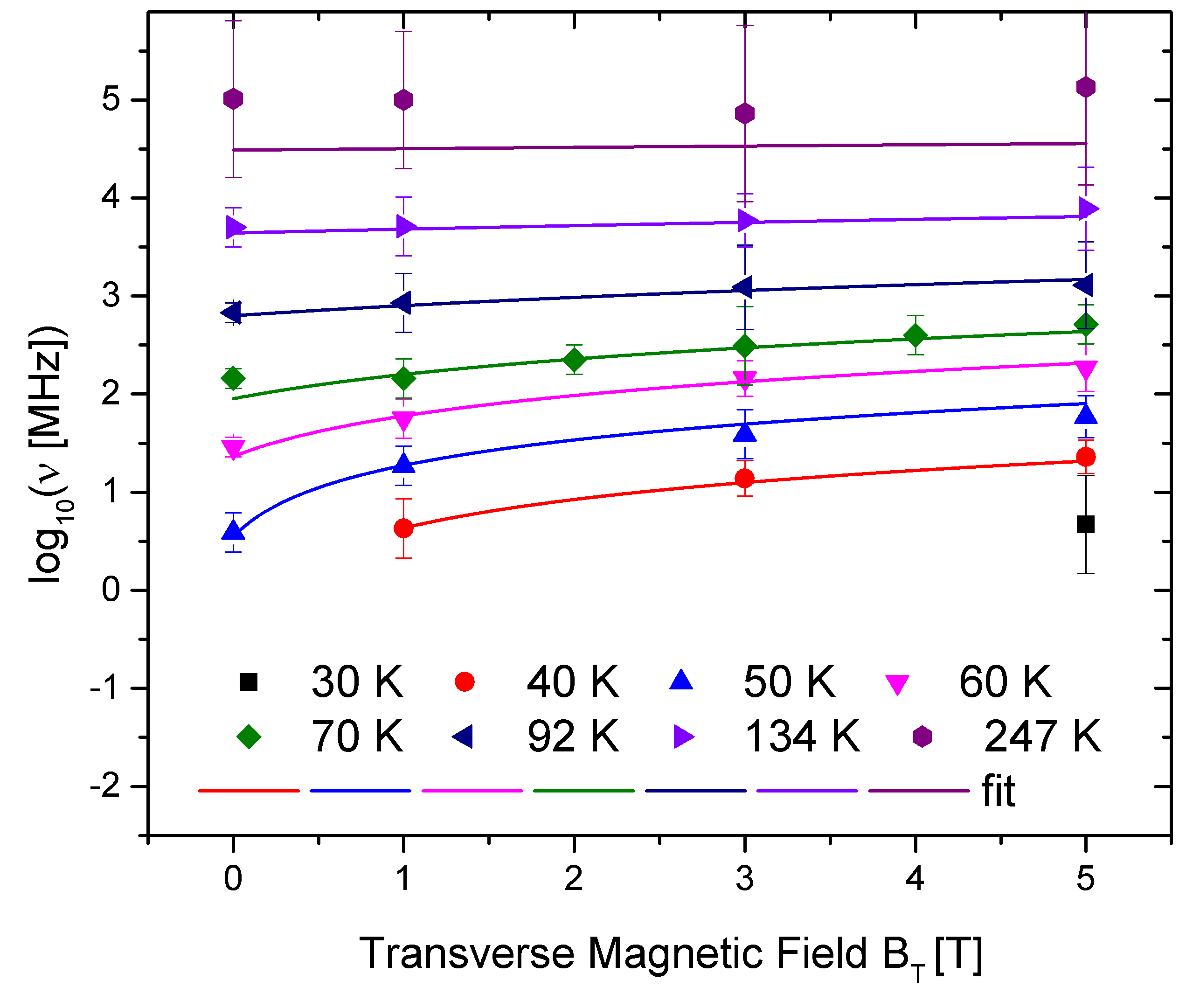}
\caption{\label{fig:Wfeld}Dependence of the spin fluctuation rate on the transverse magnetic field for \Li with $x=2.75\%$, SC 1, for various temperatures. The solid line fitting function is described in the text.}
\end{figure}

At $T = 70$\,K, $\nu$ has increased in 1\,T by a factor 2 and in 5\,T by a factor 8. This documents a strong transverse field sensitivity.
The dependence of the fluctuation frequency $\nu$ on the transverse magnetic field $B_T$ and temperature $T$ is investigated in detail for SC 1. Fig.~\ref{fig:Wtemperatur} shows the logarithmic frequency $\log_{10}$ as a function of the inverse temperature $1/T$ for different transverse external fields and Fig.~\ref{fig:Wfeld} shows the logarithm of $\nu$ as a function of the transverse magnetic field for different temperatures.\\The used fit function is discussed in the spin-Hamiltonian part below and in the appendix. In Fig.\,\ref{fig:Wtemperatur}, at low temperatures $1/T > 0.01$\,K$^{-1}$, a pronounced field-induced non-linear deviation from the zero-field Arrhenius line is observed. 
For high temperatures $1/T<0.01$\,K$^{-1}$ the data converge to the zero-field Arrhenius line, i.e. the temperature-induced fluctuations are dominant.
This is also seen in Fig.~\ref{fig:Wfeld}: the change of $\nu$ with increasing $B_T$ is enhanced by lowering the temperature. Note that for the lowest temperatures (30\,K and 40\,K) the determined fluctuations rates are close to the lower bound of the frequency window of the M\"ossbauer method due to the effective time window.


To describe the change of the spin fluctuation rates induced by the applied transverse magnetic field  we considered a simplified perturbation proportional to $B_T \, J_x \equiv B_T \, (J_++J_-)/2$ . Such a term can mix the states with different $J_z$, however with one difference from the $O_6^6$ processes discussed in section \ref{TFanalysis} below, namely $\Delta J_z=1$. We have described our experimentally observed fluctuation rate data by a function linear in $B_T$. The data can be described with the phenomenological model function
\begin{equation}
\nu(T,B_T)=\nu_0e^{-\frac{\Delta_1}{T}}+\xi B_Te^{-\frac{\Delta_2}{T}}.
\end{equation}
The first term describes the field-independent temperature-activated Arrhenius-contribution observed in the ZF experiments using $\bar{\nu}_0$ and $\Delta_1=\bar{E}_A/k_B$ 
(see Fig.~\ref{fig:row} and black line in Fig.~\ref{fig:Wtemperatur}). The second term describes the increase of $\nu$ due to the transverse field scaling linear 
with $B_T$. In a global fit $\bar{\nu}_0$, $\xi$ and $\Delta_i$ are constant parameters. The result is $\xi=2332(995)$\,MHz/T and $\Delta_2=251(20)$\,K. We associate this relaxation process with a second Orbach process which is observed by ac susceptibility below 30K (see Fig. 13). The applied transverse field increases the attempt frequency so that it becomes detectable  within the M\"ossbauer frequency window.

\subsection{$^{57}$Fe-M\"ossbauer Spectroscopy in longitudinal magnetic Fields $B_L$}
$^{57}$Fe-M\"ossbauer spectroscopy measurements were performed with applied longitudinal magnetic fields (LF) at 100\,K up to 3\,T with the $\gamma$-beam parallel to the applied field parallel to the c-axis of the crystal.
\begin{figure}
\includegraphics[width=\columnwidth]{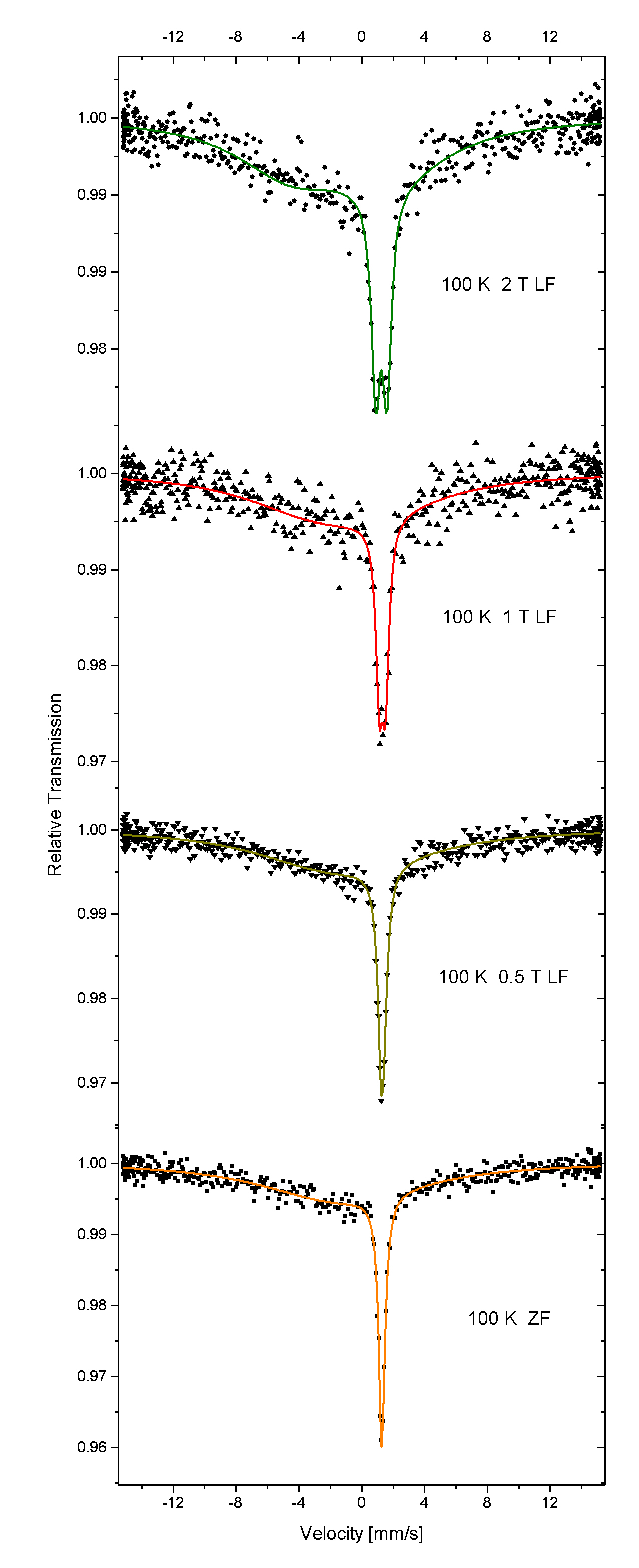}
\caption{\label{fig:LFseries}Comparison of the M\"ossbauer measurements of SC 1 at 100\,K in ZF and in applied longitudinal magnetic fields $B_L$ of 0.5\,T, 1\,T and 2\,T.}
\end{figure}
\begin{figure}
\includegraphics[width=\columnwidth]{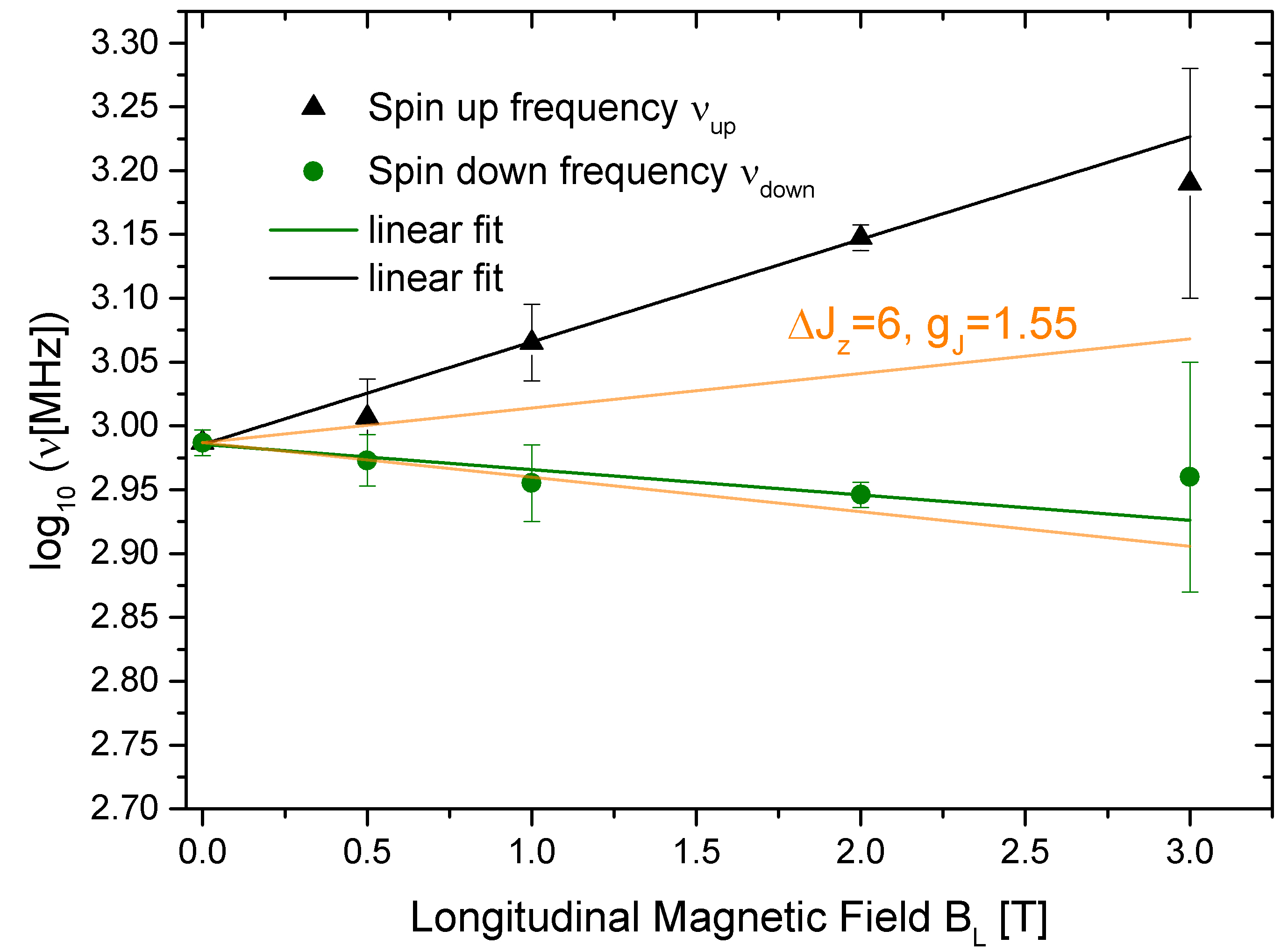}
\caption{\label{fig:LFfrequ}The fluctuation frequencies $\nu_\mathrm{up}$ and $\nu_\mathrm{down}$ as a function of the applied longitudinal magnetic field $B_L$. The dashed orange lines describe the expected values of $\nu_\mathrm{up} (B_L)$ and $\nu_\mathrm{down} (B_L)$ for Zeeman contributions to the energy levels of the $J=7/2$ Fe ions and $\Delta J_z = \pm 6$, for details see subsection III. G.}
\end{figure}

Fig.~\ref{fig:LFseries} shows the $^{57}$Fe-M\"ossbauer measurements at 100\,K up to 2\,T longitudinal magnetic field (LF). The measurements at 0.5\,T and 1\,T show an increase of the linewidth of the central absorption line compared to the ZF spectrum. The spectra at 2\,T clearly reveals a splitting into two lines corresponding to two different fluctuation rates.
The analysis model to describe the LF spectra is the Blume two-state spin reversal fluctuation model between the states with hyperfine fields $B_h= B_A+B_L$ and $B_h=- B_A+B_L$. Since the Zeeman interaction will lift the degeneracy between the "spin up" and "spin down" transitions two different fluctuation frequencies $\nu_\mathrm{up}$ describing the frequency to flip the spin into longitudinal magnetic field direction and $\nu_\mathrm{down}$ to flip it against the applied field direction are considered. The population of the two states are assumed to be the same as shown by the equal central line intensities at 2\,T in Fig. \ref{fig:LFseries}. Note that a small static external field at the $^{57}$Co-source  caused by the Helmholtz magnet leads to a slight increase of the linewidth (0.24(2)\,mm/s at 2\,T). 
Fig.~\ref{fig:LFfrequ} shows the deduced  frequencies $\log_{10}\nu_\mathrm{up}$ and $\log_{10}\nu_\mathrm{down}$ as a function of the longitudinal magnetic field $B_L$. The observed change of the fluctuation rate is one order of magnitude smaller than in the case of applied transverse fields. The data show a linear dependence of $\log_{10}\nu_\mathrm{up}$ and $\log_{10}\nu_\mathrm{down}$ as a function of $B_L$ up to 3\,T. We clearly observe an asymmetry of the observed positive and negative frequency changes, i.e. a stronger increase of $\nu_\mathrm{up}(B_L)$
than decrease of $\nu_\mathrm{down}(B_L)$. This  cannot be explained by the Zeeman-induced decrease of the energy differences for the transition $J_z = -7/2\rightarrow J_z=5/2$ and increase of the energy difference for $J_z=7/2\rightarrow J_z = -5/2$ since these changes are of equal absolute value. The experimental slopes are given by $0.084(5)$\,$\log_{10}$[MHz]/T for $\nu_\mathrm{up}$ and $-0.015(3)$\,$\log_{10}$[MHz]/T for $\nu_\mathrm{up}$, respectively. The theory curves (orange lines) shown in Fig.~\ref{fig:LFfrequ} will be discussed in the  section \ref{LFanalysis}.

\subsection{Effective single-ion $J_\mathrm{eff}=7/2$-Hamiltonian calculation of spin dynamics} \label{TFanalysis}

A striking result of the temperature and transverse magnetic field dependent M\"ossbauer spectroscopy is that  the activation energy scale for thermal fluctuations of the individual electronic Fe spins $E_A \approx 570$\,K is two orders of magnitude larger than the Zeeman energy $g_t\mu_B B_x S_x\sim5$\,K which is needed to induce similar changes of the fluctuation rate.  
Moreover it is important to note, that the low temperature longitudinal magnetic field magnetization data on this system presented in Fig.\,\ref{fig:m-h} also reveal an energy scale for longitudinal magnetic field induced system changes of the order of 1 to 5\,K from the appearance of level crossing induced magnetization steps at $\mu_0H_z = 0.15, 0.55$, and 3\,T.

For a qualitative understanding of the transverse magnetic field and temperature dependence of the spin fluctuation frequency $\nu$ we present a calculation of the spin dynamics using a single-ion spin Hamiltonian model. 
We demonstrate that an axial anisotropy of energy scale $D\ll 570$\,K, consistent with the Zeeman response of the system, can indeed give rise to an effectively two orders of magnitude larger energy barrier for thermal fluctuations. Moreover, qualitatively, the obtained results can be extended to a broad class of SAM and SMM by the introduced effective mixing term.

We consider the single-ion properties of Fe ions in Li$_2$(Li$_{1-x}$Fe$_x$)N. 
Considering spin-orbit interaction and the hexagonal point symmetry of the Fe site (1$b$ Wyckoff site, point symmetry $6/mmm$\,\cite{Klatyk1999}), the crystalline electric field yields the single-ion magnetic anisotropy  ${\cal H}_{an} = B_2^0O_2^0 +B_4^0O_4^0 +B_6^0O_6^0 +B_6^6O_6^6$, where $O_n^m$ are Stevens' operators, and $B_n^m$ are the parameters of the magnetic anisotropy\,\cite{SEGAL1970347,SEGAL197399,SEGAL1974203}. 
The Fe ions in Li$_3$N can be either Fe$^{+}$ (which, according to Hund's rules have the lowest multiplet with $S=3/2$, $L=3$ and $J=9/2$), or Fe$^{2+}$ (with $S=2$, $L=2$, and $J=4$). 
For diluted Fe in $\alpha$-Li$_3$N, however, we can apply the arguments used in Refs.\,\onlinecite{Klatyk2002, PhysRevB.66.184434}, where a ground state of $\ket{M_L=-2,M_S=-3/2}$ was proposed. The oxidation state Fe$^{1+}$ $d^7$ is consistent with our obtained hyperfine parameters, see discussion.
The arguments, namely, the strong uniaxial anisotropy of the hexagonal lattice (due to $O_2^0$, $O_4^0$ and $O_6^0$ operators, which distinguish only $z\parallel c$ axis) lifts the degeneracy, and only $L=2$ states are coupled with $S=3/2$ for Fe$^+$ in $\alpha$-Li$_3$N. 
It yields the effective total moment $J_\mathrm{eff} = 7/2$.
The splitting of the M\"ossbauer lines (see Fig.~\ref{fig:LFseries}) also confirms that assumption: The splitting is too large for pure spin states $S=3/2$ or $S=2$. In what follows we call $J$ a pseudo-spin and drop the index ``eff'' for simplicity. However, qualitatively, the calculated results below are independent of the detailed value of $J$.

The most important part of the Hamiltonian ${\cal H}_{an}$  can be written as  ${\cal H}_{an} \approx -DJ_z^2$, where $D$ is the parameter of the magnetic anisotropy.  We can conclude from the magnetization experiments and the spin-reversal Blume model with axis c\,$||B_A||V_{zz}$, that we deal with ``easy-axis'' magnetic anisotropy, $D > 0$. Consider the Hamiltonian of the Fe ion in the external magnetic field $B_z=B_L$, directed along the ``easy'' axis, namely ${\cal H}_0 = -DJ_z^2 -g_l\mu_BB_zJ_z$, where $\mu_B$ is the Bohr magneton, and $g_l$ is the (longitudinal) $z$-component of the effective $g$-tensor. The levels of that Hamiltonian cross each other at several values of $B_z$, depending on the  value. The only Stevens' operator from ${\cal H}_{an}$, which does not commute with $J_z$, and, hence, which can mix states with different values of $J_z$ and lift the degeneracies at the crossover points, is $O_6^6 \equiv (J_+^6 +J_-^6)$, where $J_{\pm} =J_x\pm iJ_y$. Such a mixing is the crucial point for the quantum tunneling~\cite{ULYANOV1992179}. Notice that according to standard quantum mechanics in the basis with diagonal action of the operator $J_z$ the eigenstates of $O_6^6$ for $J=3/2$ and $J=2$ are zero. The operator $O_6^6$ corresponds to the processes with $\Delta J_z=6$, hence connecting the states $J_z = -7/2$ with $J_z=5/2$, and $J_z = -5/2$ with $J_z=7/2$. 

Unfortunately, the explicit results for the relaxation rate due to quantum tunneling cannot be realized for $O_6^6$ because of the numerical effort. 
Therefore, to mimic the action of the operator $O_6^6$ we consider a more simplified perturbation related to the transverse magnetic field, for example, $J_x \equiv (J_++J_-)/2$. Such a term also can mix the states with different $J_z$, however with one difference from the $O_6^6$ processes, namely  $\Delta J_z=1$. This substitution of $O_6^6$ by $J_x$, while giving the opportunity to obtain a qualitative agreement with the results of our experiments, still cannot give a full quantitative description of Li$_2$(Li$_{1-x}$Fe$_x$)N. 

Summarizing, we consider, an effective Hamiltonian, which permits quantum tunneling in Li$_2$(Li$_{1-x}$Fe$_x$)N. It has the form
\begin{equation}
{\cal H} = -DJ_z^2 -g_{t}t\mu_BB_xJ_x - g_l\mu_BB_zJ_z,
\end{equation}
where $g_{t}$ is the value of the effective $g$-tensor in the plane, transverse to the easy axis. 
Note that $B_x$ can include not only the effective field, introduced to mimic the action of the $O_6^6$, but also internal (dipole) or external magnetic fields applied transverse to the $z$ axis, i.e. perpendicular to the crystallographic c axis. 
According to \cite{Zaslavskii90,ULYANOV1992179} the lowest $(2J+1)$ eigenvalues and eigenfunctions of that Hamiltonian coincide with those of the discrete spectrum of a quantum particle in the effective potential
\begin{equation}
U= \frac{DB^2}{4}\left[\sinh (x) -\frac{C}{B}\right]^2-\frac {DB}{2} (2J+1)\cosh (x),
\end{equation}
where $B=g_t\mu_BB_x/D$ and $C=g_l\mu_B B_z/D$. 
The spin quantum tunneling in that approach is totally equivalent to the tunneling of that quantum particle between the minima of the potential $U$. The tunneling rate can be calculated using the Euclidean version of dynamical equations, using dynamics of instantons of the Eulcidean action, i.e., solitons, connecting two minima of the potential $U$ with each other~\cite{0022-3719-19-30-005}. Consider the range of the field values, limited by the region $[B^{2/3} +C^{2/3}]^{3/2} < 2J+1$, in which the potential $U$ has two minima (the lowest minimum is related to the stable state, and the highest one to the metastable state). The energy barrier between the minima is finite, hence there exists a probability for the metastable state to decay due to the quantum tunneling. It is possible to calculate the values of the relaxation rate due to the quantum tunneling \cite{Zaslavskii90,ULYANOV1992179}, expanding the expression for $U$ near the position of the metastable minimum. The decay rate is determined by the analytic continuation of the energy value to the complex plane. 
Analyzing the results obtained this way, we conclude that two regimes, $T_0 \le T \le T_1$, and $T \ge T_1$, where $T_0 =  D\sqrt{a}/\pi k_B$, and $T_1 = 2D\sqrt{a}/\pi k_B$ can be related to the conditions of our experiments with Li$_2$(Li$_{1-x}$Fe$_x$)N. 
Here and below we use the notations
\begin{eqnarray}
&&a= \frac {3^{1/2}}{2^{3/2}} BC^{1/3}(2J+1)^{2/3} \delta^{1/2} \ , \nonumber \\
&&b=\frac{1}{12} B^{4/3} C^{1/3}(2J+1)^{1/3} \ , \nonumber \\
&&\delta = 1 - \frac {\left( B^{2/3} +C^{2/3}\right)^{3/2}}{2J+1}\ .
\end{eqnarray}
For $T_0 \le T \le T_1$, i.e., at low temperatures for our experiment, the relaxation rate can be approximated as, according to \cite{ULYANOV1992179},
\begin{eqnarray}
&&\hbar \gamma_1 = D \frac {\sqrt{a}\sinh(D\sqrt{a}/k_B T)}{\pi \sin (D\sqrt{a}/k_B T)} \times \nonumber \\
&&\times \exp\left[ -4Da^3/27b^2k_B T\right]\ .
\end{eqnarray}
On the other hand, for higher temperatures $T \ge T_1$ the relaxation rate is
\begin{equation}
\hbar \gamma_2 = D \frac{\sqrt{a}}{\pi} \exp\left[ -4Da^3/27b^2k_BT\right].
\label{eq-gamma2}
\end{equation}
This higher temperature behavior of the relaxation rate caused by the quantum spin tunneling is similar to the Orbach relaxation \cite{Orbach61}, i.e., it has the Arrhenius form. 
Notice that the ``true'' quantum spin tunneling-induced relaxation rate exists only at $T = 0$ \cite{Zaslavskii90,ULYANOV1992179}.

Given that the pre-expontial factor, $\nu_0$, obtained from M\"ossbauer spectroscopy is temperature-independent (as the pre-exponential factor in equation\,\ref{eq-gamma2}), the most essential regime for our experiments with Li$_2$(Li$_{1-x}$Fe$_x$)N is the region with $T \ge T_1$. 
We see that the relaxation rate $\gamma_2$ follows an Arrhenius law in the temperature dependence, $\gamma_2 = \nu_0 \exp(-E_A/k_BT)$ with the prefactor $\nu_0$ and the activation energy $E_A$ determined as
\begin{eqnarray}
&&\nu_0 = \frac {D\sqrt{a}}{\pi \hbar} \sim B_x^{1/2} B_z^{1/6} \delta^{1/4} \ , \nonumber \\
&&E_A = \frac {4Da^3}{27b^2} \sim (B_xB_z)^{1/3} \delta^{3/2} \ .
\end{eqnarray}
In detail, 
\begin{equation}
E_A\approx 4.89B^{1/3}C^{1/3}(2J+1)^{4/3}\delta^{3/2}
\end{equation}
where the factor $(2J+1)^{4/3}$ is the dominant scaling factor for $J>3/2$. 

\begin{figure}
\begin{center}
\vspace{-15pt}
\includegraphics[width=\columnwidth]{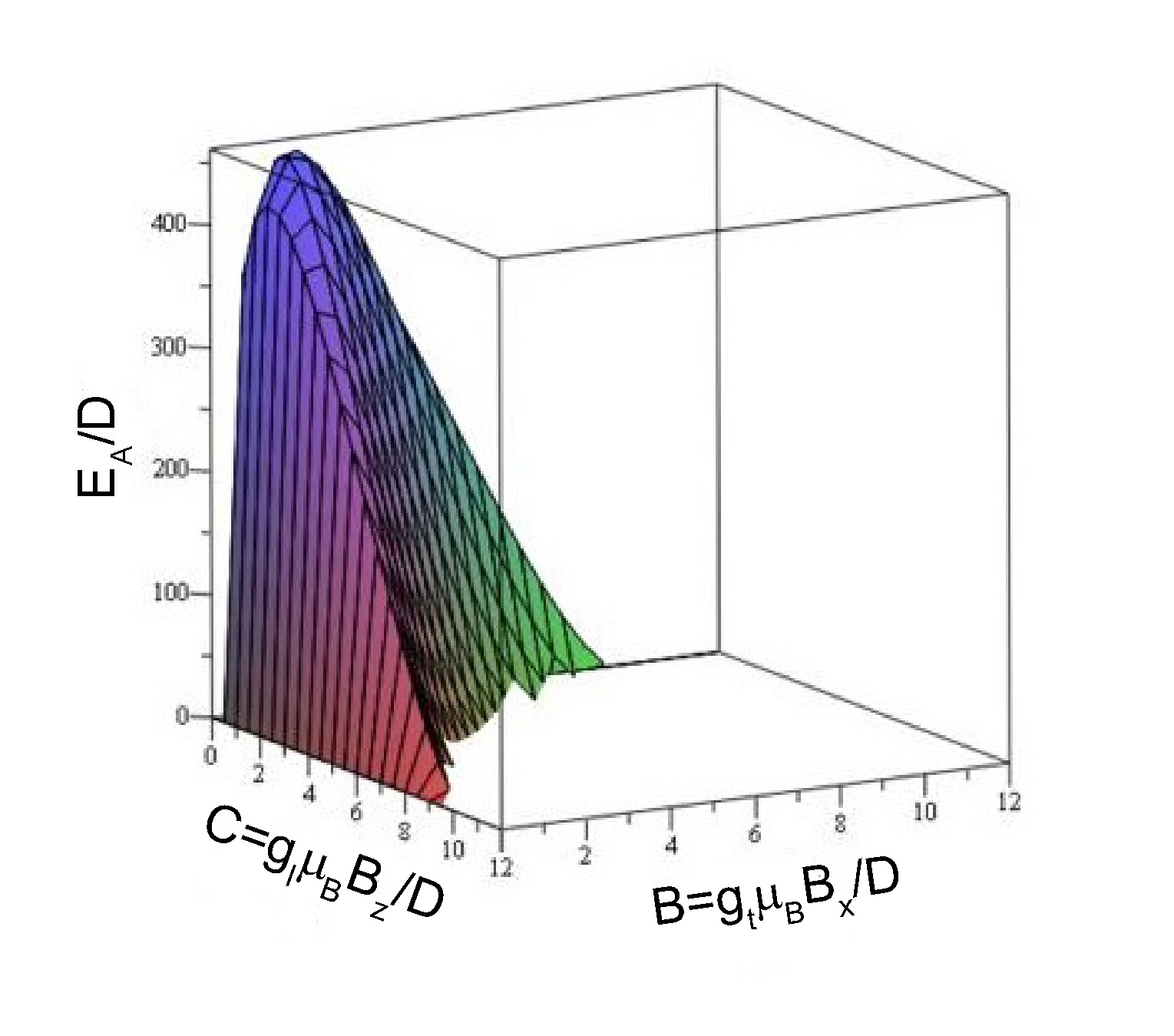}
\end{center}
\vspace{-15pt}
\caption{The dependence of the activation energy $E_A$ for the relaxation rate $\gamma_2$ on the longitudinal $B_z$ and transverse $B_x$.}
\label{fig1}
\end{figure}

Note that we cannot use the limits $B_x \to 0$, and $B_z \to 0$ in the expression for the relaxation rate caused by the spin tunneling, because the latter is absent there: Without $B_z$ there is no crossover, and without $B_x$ there is no lifting of the degeneracy of crossover points). However, we can evaluate the field dependence of the activation energy, not taking into account the limiting cases $B_x \to 0$ and $B_z \to 0$.
A weak effective tilted magnetic field can  originate, e.g., from the long-range magnetic dipole-dipole interaction in the mean field approximation. 
We also suppose that the region of applicability of the results can be expanded to all $\delta < 1$, which implies the  difference between the potential $U$ and its expansion near the position of the metastable state being small (this difference produces higher-order quantum corrections). 
The result is shown in Fig.\,\ref{fig1} for $J=7/2$.
We see that for very small but finite values of the components of the external magnetic field the activation energy is much larger than the value of the magnetic anisotropy $D$. 
It explains the observation of the giant activation energy for the relaxation rate in our M\"ossbauer studies of Li$_2$(Li$_{1-x}$Fe$_x$)N. 
Furthermore, we see that the application of the external field of the order of $(2J+1)D$ reduces drastically the value of the activation energy. 

\begin{figure}
\begin{center}
\vspace{-15pt}
\includegraphics[width=\columnwidth]{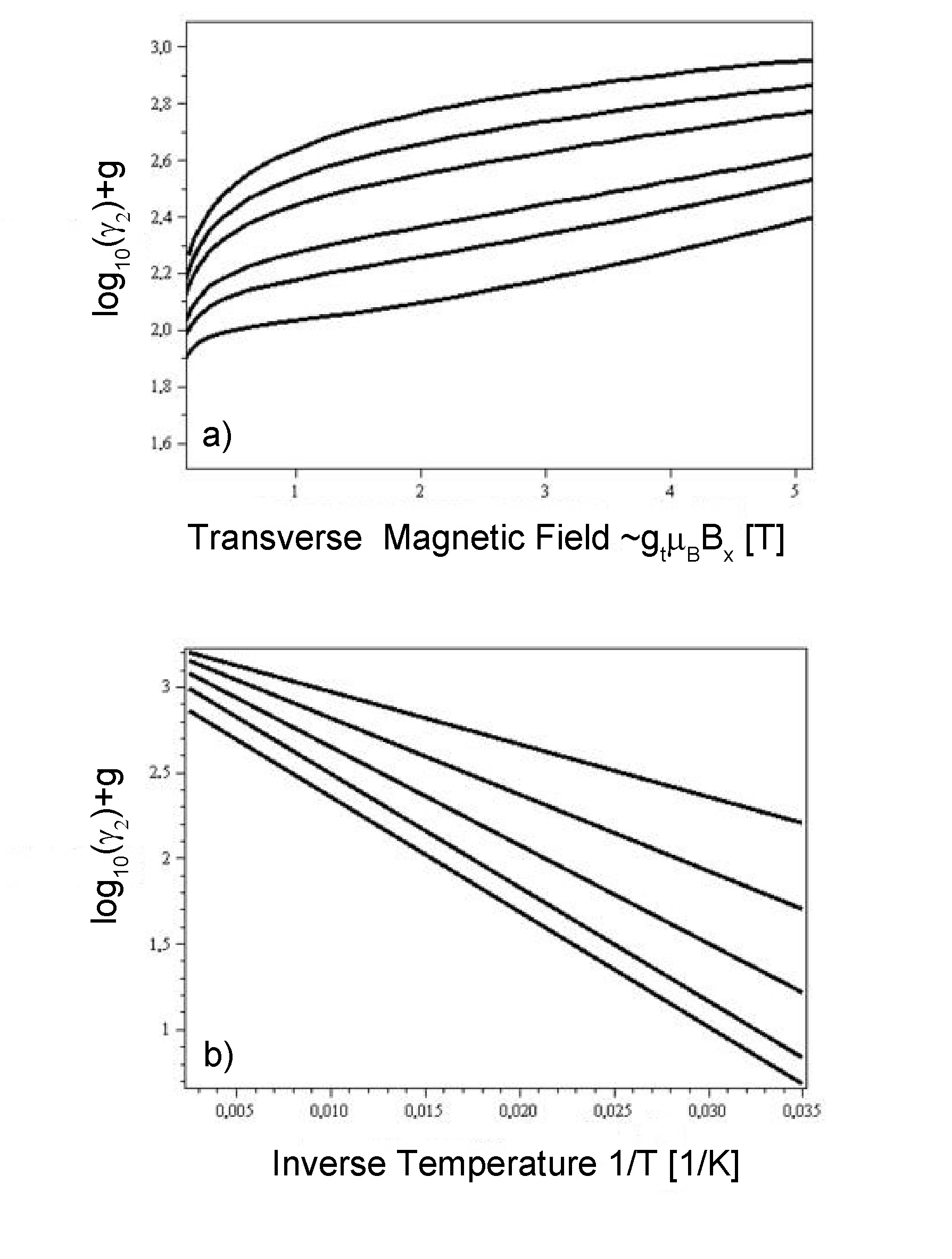}
\end{center}
\vspace{-15pt}
\caption{a) The dependence of the logarithm of the relaxation rate $\gamma_2$ on the transverse field $B_x$ for several values of $T$. From top to bottom: $T=247$~K, $T=134$~K, $T=92$~K, $T=60$~K, $T=50$~K and $T=40$~K. b) The dependence of the logarithm of the relaxation rate $\gamma_2$ on the inverse temperature 1/$T$ for several values of the transverse field $B_x$. From bottom to top: $B_x=1$\,T, $B_x=2$\,T, $B_x=3$\,T, $B_x=4$\,T $B_x=5$\,T.}
\label{fig2}
\end{figure}

Now we can compare the transverse field dependence of the relaxation rate, extracted from M\"ossbauer experiments in Li$_2$(Li$_{1-x}$Fe$_x$)N with the calculated one. 
In Fig.\,\ref{fig2} a) the logarithm of the relaxation rate $\gamma_2$ is plotted as a function of the applied transverse field $B_x$ at $g_l\mu_BB_z=0.001$ for $J=7/2$ and several values of the temperature. 
To have better agreement with experiment we have to add the constant $g=3.5$ to $\log_{10}(\gamma_2)$, which implies additional sources of relaxation that are temperature- and magnetic field-independent.
 Fig.\,\ref{fig2} b) shows the logarithm of the relaxation rate $\gamma_2$  as a function of the inverse temperature, $1/T$, for several values of the transverse external magnetic field $B_x$.
We see that the general tendency is well described by our simplified theory, while there is no quantitative agreement. 

We conclude that this single-ion theory, based on the spin properties of Fe impurities, which at low energies produce quantum spin tunneling, well reproduces the most dramatic feature of dynamical experiments in Li$_2$(Li$_{1-x}$Fe$_x$)N: the giant value of the activation energy in the Arrhenius law for the temperature dependence of the relaxation rate, and much smaller values of the external magnetic field, which drastically change that relaxation rate. 

\subsection{Zeeman Analysis of Spin Dynamics in longitudinal Fields $B_L$} \label{LFanalysis}

\begin{figure}[h]
\includegraphics[width=\columnwidth]{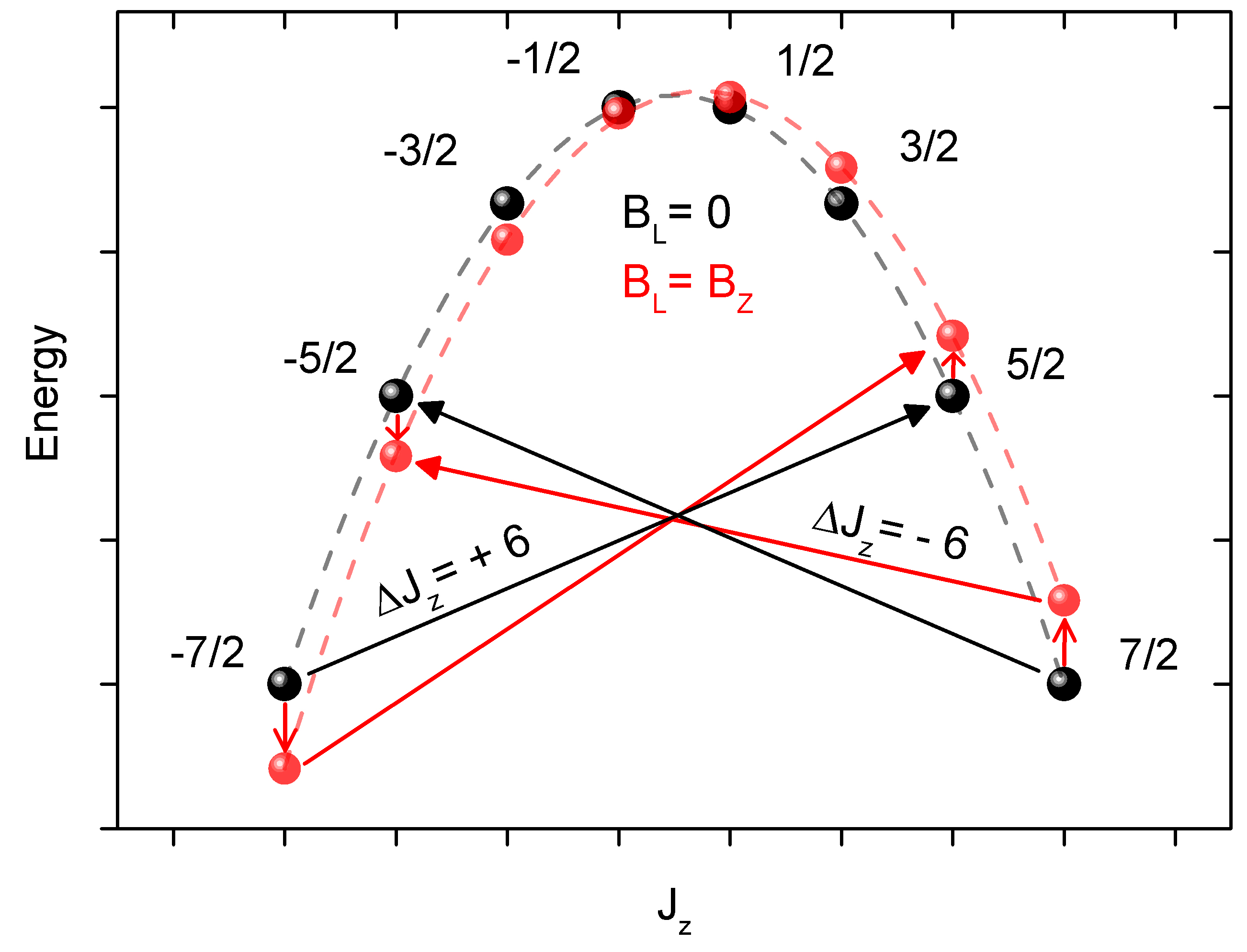}
\caption{\label{fig:transitions}Illustration of the transitions between $J_z = -7/2$ and $J_z=5/2$ and $-7/2$ to $+5/2$. The energy levels $J_z$ are shifted in LF (red points) leading to different transition energies according to the Zeeman energy $E_Z \sim J_z \, B_Z$.}
\end{figure}

The splitting of the  resonance line in longitudinal fields (Fig.\,\ref{fig:LFseries}) can be understood as a consequence of the Zeeman term in the effective spin Hamiltonian. 
For $B_z=0$ the relevant relaxation processes with $\Delta J_z=\pm 6$ introduced in the last section are equivalent. However, a finite longitudinal field $B_z=B_L$ removes the degeneracy of the $\pm J_z$ energy levels via Zeeman interaction. 
The values of the energy differences between the states $J_z = -7/2$ and $J_z=5/2$, and between the states $J_z = -5/2$ and $J_z=7/2$ become non-equal (see Fig.\, \ref{fig:transitions}).

To calculate these energy differences we need to specify the longitudinal g-factor 
$g_l$. Here, we model the complex many electron state of the Fe ions including strong spin-orbit interaction  by a simplified  $J = 7/2$ system with a fixed Land\'e-g-factor value of $g_l = 1.55$ (based on $L=2$ and $S=3/2$, see appendix) for both, the ground state (assumed to be $J_z = \pm 7/2$) and the excited state (assumed to be $J_z = \pm 5/2$). This choice of $g_l$ leads to an effective paramagnetic moment $\mu_\mathrm{eff}=g_l\,\mu_B\sqrt{J(J+1)}\approx 6.15 \, \mu_B$ which is consistent with the experimental value.


According to the Arrhenius law the change 
of the spin fluctuation rate with respect to its ZF value can be calculated according to 
\begin{equation}
\nu (B_L) / \nu (B_L = 0) = \exp{\left(\frac{E_Z}{k T}\right)}.
\end{equation}
Here, $E_Z=\pm g_l\mu_B \Delta J_z B_L$ is the Zeeman term with the transition rule $\pm \Delta J_z$. The $\pm$-sign corresponds to the positive and negative branches in Fig.~\ref{fig:LFfrequ}. To calculate  the Zeeman induced change of the spin fluctuation rate $\nu (B_L)$, we use $\Delta J_z = 6$ according to the hexagonal symmetry. The resulting slope of $\nu$ vs. $B_L$ is
\begin{equation}
\frac{\Delta\log_{10}\nu}{\Delta B_L}\approx\pm0.027/\mathrm{T}.\label{LFslope}
\end{equation}
The results of this calculation are included in Fig.~\ref{fig:LFfrequ} as orange lines. The calculated values for $\nu(B_L)$ are below the experimental results for the branch $\nu_\mathrm{up}(B_L)$ as well as the branch $\nu_\mathrm{down}(B_L)$. This is expected since a second contribution stemming from the $B_L$ dependence of the $B^6_6O^6_6$ term in the crystal field spin Hamiltonian is not included in this model. Such a term  is always positive, linear in $B_L$ and identical for both branches.
Therefore, the  Zeeman contributions to $\nu_\mathrm{up}$ and 
 $\nu_\mathrm{down}$ must be located below the experimental values, ideally shifted by  identical values with respect to the experimental values.
 The latter is not fulfilled (crf. Fig.~\ref{fig:LFfrequ}), however increasing the absolute value of both slopes (equation \ref{LFslope}) by $\approx$ 80\% would lead
 to such a situation. Note that a correction of this size is feasible since from all experiment performed on \Li so far we cannot  determine the ground state and excited state values of $J_z$ exactly. Moreover, also the longitudinal g-factors $g_l$ for both states in this effective spin Hamiltonian approach can be modified strongly due to the subtle interplay of the crystal electric field with the spin-orbit coupling in this 3$d^7$ state. 
 A more realistic many-body electronic structure calculation is needed to calculate the effective crystal field energies as well as the longitudinal and transverse g-factors of the ground and excited states seperately.

\section{Discussion}

\subsection{M\"ossbauer Sites and Sample Homogeneity}

Two Fe sites A and B are observed in the low temperature M\"ossbauer spectroscopy on sample SC 1. The main site A is associated with monomer Fe sites without relevant magnetic exchange with other Fe ions since it is observed also in samples SC 2-4 which contain an up to one order of magnitude lower Fe concentration $x$.
Site B is not observed in SC 2-4. We associate site B with a nearest neighbor in-plane or out-of-plane Fe-dimer site. The magnetic hyperfine field  for the two Fe subspecies is determined to $B_A=70.21(1)$\,T and $B_B=65.0(2)$\,T at 2\,K. These values are in agreement with Refs.~\cite{Klatyk2002,ZAAC:ZAAC200300135,Fix2018b}, in which Klatyk \textit{et al.}, Ksenofontov \textit{et al.} have performed a powder study of $x > 0.15$ proposing ferromagnetic ordering for $T<65$\,K.

The temperature-dependence of the M\"ossbauer spectrum shown in Fig.~\ref{fig:2} is consistent with the expected behavior of SAM. The observed spin fluctuations are consistently described by a thermal activation crossover rather than by a cooperative long-range ordering transition. However, this does not exclude by itself that Fe site A arises from small cluster-like SMM units like Fe$_i$ clusters in the Li$_3$N matrix with ferromagnetic interaction between the Fe ions with various size numbers $i$ of Fe depending on $x$. The deduced hyperfine parameters are within error bars identical for samples SC 1-4. The spin dynamics described by the fluctuation frequency $\nu (T)$, $\nu_0$ and $E_A$ are concentration-independent for Fe site A of SC 1-4. The invariant parameters as a function of $x$ proves well isolated Fe sites like in a SAM. 



A combinatorial expression to calculate the probability for n Li ions among six neighbors in the [001] plane for the Fe concentration $x$ yields
\begin{equation}
W_n=6![n!(6-n)!]^{-1}(1-x)^nx^{6-n}\approx 14\%
\end{equation}
for $x = 0.027$ and $n= 5$, i.e. an in-plane Fe-dimer~\cite{Klatyk2002}. This value is twice as large as the observed value. The area contribution of site B is overestimated in this statistical treatment in which every kind of Coulomb repulsion is neglected. Either due to Coulomb repulsion a more homogeneous mononuclear SAM is preferred or an out-of-plane Fe-N-Fe dimer configuration is the observed site B.
Interestingly, the total contribution of the Fe-N-Fe in a binomial distribution is supposed to be $\approx 5.3$\% which is closer to the experimentally determined value of 5.9(3)\% of Fe site B. 
A systematic M\"ossbauer study on a series of \Li~with larger $x$ on single-crystals is needed to identify nearest neighbor Fe-cluster configurations in this system. 
However, this is beyond the scope of this work.

\subsection{Oxidation and Spin State of Fe Ions in \Li}

The observed isomer shift value around 0.10\,mm/s is unconventional for a Fe oxidation states of Fe$^{1+}$ or Fe$^{2+}$. 
It can possibly arise from the linear N-Fe-N low-coordinated electronic structure of Fe in \Li. Because of the paramagnetic behavior a Fe$^{2+}$ low spin state $S=0$ can be excluded.

A $^{57}$Fe-M\"ossbauer study was performed on the linear complexes [K(crypt-222)][Fe(C(SiMe$_3$)$_3$)$_2$] and [Fe(C(SiMe$_3$)$_3$)$_2$] with a similar Fe linear coordination by carbon~\cite{zadrozny_mossbauer_2013}. 
[Fe(C(SiMe$_3$)$_3$)$_2$]$^{1-}$ in [K(crypt-222)][Fe(C(SiMe$_3$)$_3$)$_2$] is proposed to contain Fe$^{+1}$, whereas Fe$^{2+}$ is present in [Fe(C(SiMe$_3$)$_3$)$_2$]. 
The assumed asymmetry parameter $\eta=0$ is according to the axial symmetric EFG tensor discussed by Lewis \textit{et al.}\,\cite{Lewis:a19705} (and references therein).
\begin{table}
\begin{tabular}{c|c|c|c}
Compound&$\delta$ [mm/s]&$\Delta E_Q$ [mm/s]&$B$ [T]\\
\hline
[Fe(C(SiMe$_3$)$_3$)$_2$]$^{1-}$		&0.402(1)	&-2.555(2)	&63.68(2)\\
{[Fe(C(SiMe$_3$)$_3$)$_2$]}				&0.460(3)	&-1.275(5)	&150.7(1)\\
Li$_2$(Li$_{1-x}$Fe$_x$)N $(x\ll 1)$ 	&0.100(2)	&-2.572(2)	&70.25(2)\\
\hline
\end{tabular}
\caption{Characteristic M\"ossbauer parameters obtained at $T \sim 4$\,K assuming $\eta=0$ compared with Zadrozny \textit{et al.}~\cite{zadrozny_mossbauer_2013}.}\label{table3}
\end{table}
Table~\ref{table3} shows the values of the isomer shift $\delta$, the quadrupole splitting $\Delta E_Q$ and the magnetic hyperfine field $B$. 
The smaller $\delta$-value of Fe site A can be explained by the increase of $s$-electron density at the nucleus and the $4s$-mixing. 
The EFG value $V_{zz}$, here given by $\Delta E_Q$ of Fe-diluted Li$_2$(Li$_{1-x}$Fe$_x$)N (site A) and the Fe$^{1+}$-SMM, $[$Fe(C(SiMe$_3$)$_3$)$_2$$]$$^{1-}$, are very close to each other whereas the Fe$^{2+}$-SMM shows only half of this value. 
Moreover, also the magnetic hyperfine fields of $[$Fe(C(SiMe$_3$)$_3$)$_2$$]$$^{1-}$ and Fe-diluted Li$_2$(Li$_{1-x}$Fe$_x$)N are comparable. 
Therefore, we conclude a strong similarity of the Fe electronic systems in these two systems with oxidation state Fe$^{+1}$ for \Li. An oxidation state of Fe$^{1+}$ is also consistent with calculated electronic band structure\,\cite{Klatyk2002,PhysRevB.66.184434}.

\subsection{Energy Barrier and Spin Dynamics}

\begin{figure}
\includegraphics[width=\columnwidth]{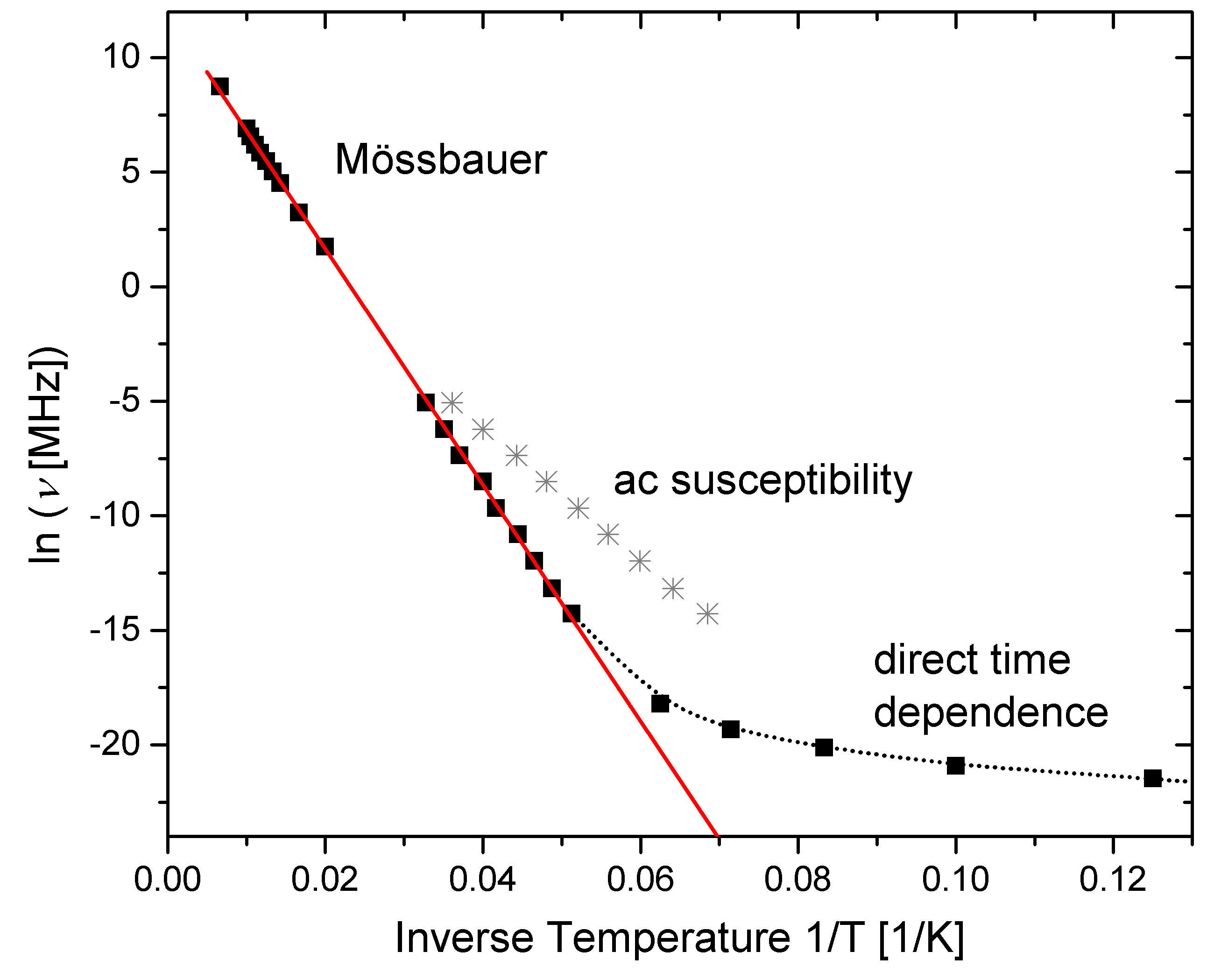}
\caption{\label{fig:rates} Fluctuation frequencies of SC 1 determined by M\"ossbauer spectroscopy, ac susceptibility and  magnetization relaxation measurements (direct time dependence), shown in form of an Arrhenius plot. The red line shows a fit to the high temperature region and yields an effective energy barrier of $\bar{E}_A = 515(20)$\,K. The gray points describe a second low frequency activated relaxation process with an activation energy of $\bar{E}_A = 270(12)$\,K , observed in the ac-susceptibility at low temperatures.}
\end{figure}

Fig.~\ref{fig:rates} shows the temperature-dependence of the spin fluctuation rate of SC 1, determined from M\"ossbauer spectroscopy, ac-susceptibility  and direct magnetization relaxation measurements, respectively. 
At three M\"ossbauer data points at $1/T \gg $ 0.05\,K$^{-1}$ essentially static M\"ossbauer spectra are measured, i.e. the fluctuation rate becomes smaller than the lower bound of the frequency window of the method and these data points are not shown in Fig.\,\ref{fig:rates}.

\begin{table}
\begin{tabular}{c|c|c|c}
Compound					& magn. unit	& $E_A$ [K]	&	Reference							\\
\hline
$[$Dy(bbpen)X$]$			& Dy$^{3+}$			& 1025		& \onlinecite{doi:10.1021/jacs.6b02638}	\\
TbBis(phthalocyaninate)		& Tb$^{3+}$			& 940		& \onlinecite{Ganivet2013}				\\
Li$_2$(Li$_{1-x}$Fe$_x$)N	& Fe$^{1+}$			& 570(6)	& this work							\\
$[$Fe(C(SiMe$_3$)$_3$)$_2$$]$$^{1-}$&Fe$^{1+}$	& 354		&\onlinecite{zadrozny_mossbauer_2013, zadrozny_magnetic_2013}   \\
$[$Sr$_{10}$(PO$_4$)$_6$(Cu$_x$OH$_{1-x-y}$)$_2]$&Cu$^{3+}$&69&\onlinecite{doi:10.1002/chem.201303136}\\\hline
\end{tabular}
\caption{Selected SAMs with a large energy barrier $E_A$ and the corresponding paramagnetic ion.}\label{table4}
\end{table}

The relaxation rates, $\nu$, obtained by M\"ossbauer and ac susceptibility measurements can be well described by a single effective energy barrier of $\bar{E}_A=515(20)$\,K.
Note that there is another, larger peak observable in the temperature-dependent ac susceptibility that corresponds to a faster relaxation with an effective energy barrier of $\bar{E}_A = 270(12)$\,K (grey stars in Fig.\,\ref{fig:rates}).
The pre-exponential factor of the Arrhenius behavior amounts to $\nu_0 = 150(10)$\,GHz.
We conclude that this relaxation process is enhanced  by the application of transverse magnetic fields and becomes visible in the M\"ossbauer frequency window (see section \ref{MinBtrans} above). The deduced activation energies are consistent within error bars.
For $T < 10$\,K the relaxation rates were determined by fitting the time-dependent magnetization to stretched exponential $M(t) = M_0 \exp[-(t\,\nu)^{\beta}]$.

At zero external field the two allowed phonon-assisted relaxation processes $J_z = -7/2$ $\rightarrow$  $J_z=5/2$ and $J_z=7/2$ $\rightarrow$ $J_z = -5/2$ have equal energy differences. They become non-equal under applied longitudinal magnetic field as presented above.
The direct quantum tunneling regime  is reached below 10\,K (see Fig. \ref{fig:rates}).

In table~\ref{table4} we compare the thermal activation energy barriers for several SAM systems with large energy barriers $E_A$ compared to Fe-diluted Li$_2$(Li$_{1-x}$Fe$_x$)N. The thermal activation barrier $E_A$ is often associated with a two-phonon Orbach process~\cite{gutlich_mossbauer_2011}. Above 50\,K the dominant character of this process is plausible: the direct spin transition process in the Debye model accompanied by the creation or annihilation of a single phonon is dominant only for lower temperature $T<50$\,K with $\tau_S^{-1}\propto T$.

In the literature the energy barrier is often identified with the zero-field splitting value $D$ rather than considered as an effective experimental quantity, which depends on different microscopic parameters as discussed above. However, as demonstrated by our spin Hamiltonian approach, the energy barrier is a function of the (internal or applied) transverse magnetic field
 and a general scaling proportional to $(2J+1)^{4/3}$ (crf. eqn. (23)). The effective spin Hamiltonian calculation presented in this work can qualitatively account for the temperature and transverse field dependencies of the experimentally observed spin fluctuation rates. 

The spin dynamics in applied longitudinal magnetic fields can be understood considerung the Zeeman shift of the $J_z$ states which induces a splitting of the spin fluctuation rate into two branches. The observed experimental asymmetry is expected theoretically and is caused by higher-order Stevens' operator terms produced by the hexagonal symmetry of the lattice.

\section{Conclusions}

In this work, we present $^{57}$Fe-M\"ossbauer studies on
diluted Fe centers in a linear N-Fe-N configuration along the crystallographic c-axis in single crystalline specimen of hexagonal \Li. 
The homogeneity of the nanoscale distributed isolated Fe centers is shown and the single-atomic magnet nature confirmed. 
Below 30 K the magnetically isolated single-ion Fe centers exhibit a large quasistatic magnetic hyperfine field of $\bar{B}_A=70.25(2)$\,T parallel to the c-axis which is the strongest principle axis of the electric field gradient $\bar{V}_{zz}=-154.0(1)$\,V/\AA$^2$.

Fluctuations of the magnetic hyperfine field clearly observed in the M\"ossbauer spectra between 50\,K and 300\,K are described by a Blume two-level relaxation model. The spin dynamics in \Li is concentration-independent for $x \leq 0.028$.
 From the temperature dependence an Orbach process is deduced as the dominant spin-lattice relaxation process. An Arrhenius analysis $\nu=\nu_0e^{-E_A/k_BT}$ yields a thermal activation barrier of $\bar{E}_A=515(20)$\,K and an attempt frequency $\bar{\nu}_0=150(10)$\,GHz. M\"ossbauer spectroscopy studies with applied transverse magnetic fields up to 5\,T reveal a huge increase of the fluctuation rate by two orders of magnitude. In applied longitudinal magnetic fields a characteristic splitting of the spin fluctuation frequency is observed. These experimental observations are qualitatively reproduced by a single-ion spin Hamiltonian analysis. 
 It demonstrates that for dominant magnetic quantum tunneling a weak axial single-ion anisotropy $D$ of the order of a few Kelvin can cause a two orders of magnitude larger energy barrier for temperature-induced longitudinal spin fluctuations.
We think that this is one of the most spectacular manifestations of the macroscopic quantum spin tunneling observed in the solid-state based single-atomic magnet Li$_2$(Li$_{1-x}$Fe$_x$)N.
The experiments suggest  \Li as a candidate for novel functional magnetic materials, e.g. for quantum computing or spintronic devices.

\section{Acknowledgments}
This work was supported by the Deutsche Forschungsgemeinschaft (DFG, German Research Foundation) through SFB 1143 and JE748/1. Special thanks to J. Schnack, M. Baker and E. Bill for helpful and constructive remarks.

\section{Appendix}
\subsection{Mass absorption coefficients}
The Fe concentration of sample SC 4 is with below 0.2\% rather small, even for a $^{57}$Fe-M\"ossbauer experiment of a non-$^{57}$Fe-enriched sample. Fortunately, \Li\, contains only light elements with small absorption coefficients of the 14.41\,keV radiation energy, see table~\ref{table2}. The mass absorption coefficient $\mu_a$ describes the exponential Intensity reduction of the initial $\gamma$-ray intensity $I_0$,
\begin{equation}
I(d)=I_0(1-f_s)e^{\mu_ad},
\end{equation}
where $d$ is the absorber thickness and $f_s$ the recoil-free fraction of transitions. $\mu_a$ describes the non-resonance atomic absorption, mainly by the photoelectric effect. For a comparison, the value of Osmium represents a heavy element in table~\ref{table2} showing the rather small mass absorption coefficient $\mu_a$ of Li and N since Fe is highly-diluted. We have used for this reason large crystals of a thickness of a few millimeter and the effective thickness $t_a=2.39(10)$ of SC 1 reflects still absorption far away from the saturation limit. The Fe concentration of SC 4 is even below the concentration of Ho in LiY$_{0.998}$Ho$_{0.002}$F$_4$ or at least in the same order which is a prominent example for a SAM in a solid crystal~\cite{PhysRevLett.91.257204,PhysRevLett.87.057203}.
\begin{table}
\begin{tabular}{c|c|c}
Element  &Atomic mass [u] &absorption coefficient [cm$^2$/g]\\\hline
Li &3& 0.277\\
N  &7&1.4\\
Fe &26&64\\
Os &76&165\\\hline
\end{tabular}
\caption{Mass absorption coefficients $\mu_a$ according to Chen \textit{et al.} and therein~\cite{chen_mossbauer_2007}.}\label{table2}
\end{table}
\subsection{Magnetic Hyperfine field}
The results of the calculations are discussed assuming the Fe$^+$ oxidation state~\cite{PhysRevB.66.184434}.
In general, the total magnetic hyperfine field $B$ is the sum of different contributions
\begin{equation}
B_h=B_c+B_{\mathrm{orb}}+B_{\mathrm{dip}}+B_{\mathrm{lat}}.
\end{equation}
The sign of the Fermi contact contribution $B_c$ is negative and arises from the spin-polarization of the $s$-electrons by unpaired valence electrons. $B_{\mathrm{orb}}$ is the orbital contribution scaling with the orbital quantum number $L$ which is expected to be important because of the exceeded spin only value of the magnetic moment. $B_{\mathrm{dip}}$ is the dipolar contribution arising from nonsperical electron spin density contribution which is approximately proportional to $V_{zz}$. $B_{\mathrm{lat}}$ is the lattice contribution, i.e., the magnetic field generated by neighbor electronic moments in the lattice. This contribution can be neglected in the diluted system.   The detailed values vary strongly on the used computational method and estimations~\cite{PhysRevB.66.184434}, however, a tendency is given by
\begin{equation}
B_{\mathrm{orb}}>B_{\mathrm{dip}}\approx |B_c|>B_{\mathrm{lat}}\approx 0
\end{equation}
or even $B_{\mathrm{orb}}\gg B_{\mathrm{dip}},|B_c|$ which is based the Fe$^{1+}$ $(S=3/2$, $L=2)$ assumption~\cite{PhysRevB.66.184434}.

\subsection{Magnetization hysteresis loops}
The presented hysteresis loops of magnetization were measured at different temperatures for magnetic fields applied parallel to the crystallographic c-axis, $B \parallel c$. 
The obtained data were corrected for the diamagnetic sample holder (sample sandwiched between two torlon discs and fixed inside a straw) for which the magnetization was determined separately using a similar setup. 
The diamagnetic contribution of the $\alpha$-Li$_3$N host was subsequently subtracted from the sample holder corrected data using $\rho$(Li$^{1+})= -8.8\cdot 10^{-12}$\,m$^3$mol$^{-1}$ \,ref. \cite{Banhart1986} and $\rho$(N$^{3-})=1.63\cdot 10^{-10}$\,m$^3$mol$^{-1}$~ \,ref. \cite{Hohn2009}.

\subsection{Breit-Wigner formula}
The cross-section
\begin{equation}
\sigma(E)=\frac{\sigma_a(E)}{\sigma_0}
\end{equation}
of the transmission integral is given by the Breit-Wigner formula
\begin{equation}
\sigma_0(E)=\frac{\sigma_0\Gamma_a^2/4}{(E-E_0)^2+\Gamma_a^2/4}
\end{equation}
where
\begin{equation}
\sigma_0=\left(\frac{\lambda^2}{2\pi}\right)\frac{1+2I_{(e)}}{1+2I_{(g)}}\frac{1}{1+\alpha}
\end{equation}
is the maximum cross section, e.g see Chen~\cite{chen_mossbauer_2007}. Here, $\alpha$ is the internal conversion coefficient, $I_{(e)}$,$I_{(g)}$ are the nuclear spin numbers of the ground state and excited state, respectively, and $\lambda$ the energy of the $\gamma$-ray. $\sigma(E)$ and $\sigma_0(E)$ are given as a function of the photon energy $E$, $E_0$ is the energy of the $\gamma$-ray corresponding to the M\"ossbauer transition. The excited state is not strict monochromatic and has a natural distribution given by a Lorentzian line
\begin{equation}
L(E)dE=\frac{\Gamma_s}{2\pi}\frac{1}{(E-E_0)^2+\Gamma_s^2/4}dE
\end{equation}
with
\begin{equation}
\int L(E)dE=1.
\end{equation}
$\Gamma_s$ is the natural linewidth of the M\"ossbauer nuclei and $\Gamma_a$ is the natural linewidth of the absorber. Here,
\begin{equation}
L(E,v)=L\left(E-\frac{v}{c}E_0\right)
\end{equation}
is the relation to our notation with the speed of light $c$.
\subsection{Arrhenius plot} \label{App-Arrhenius}
\begin{figure}
\includegraphics[width=\columnwidth]{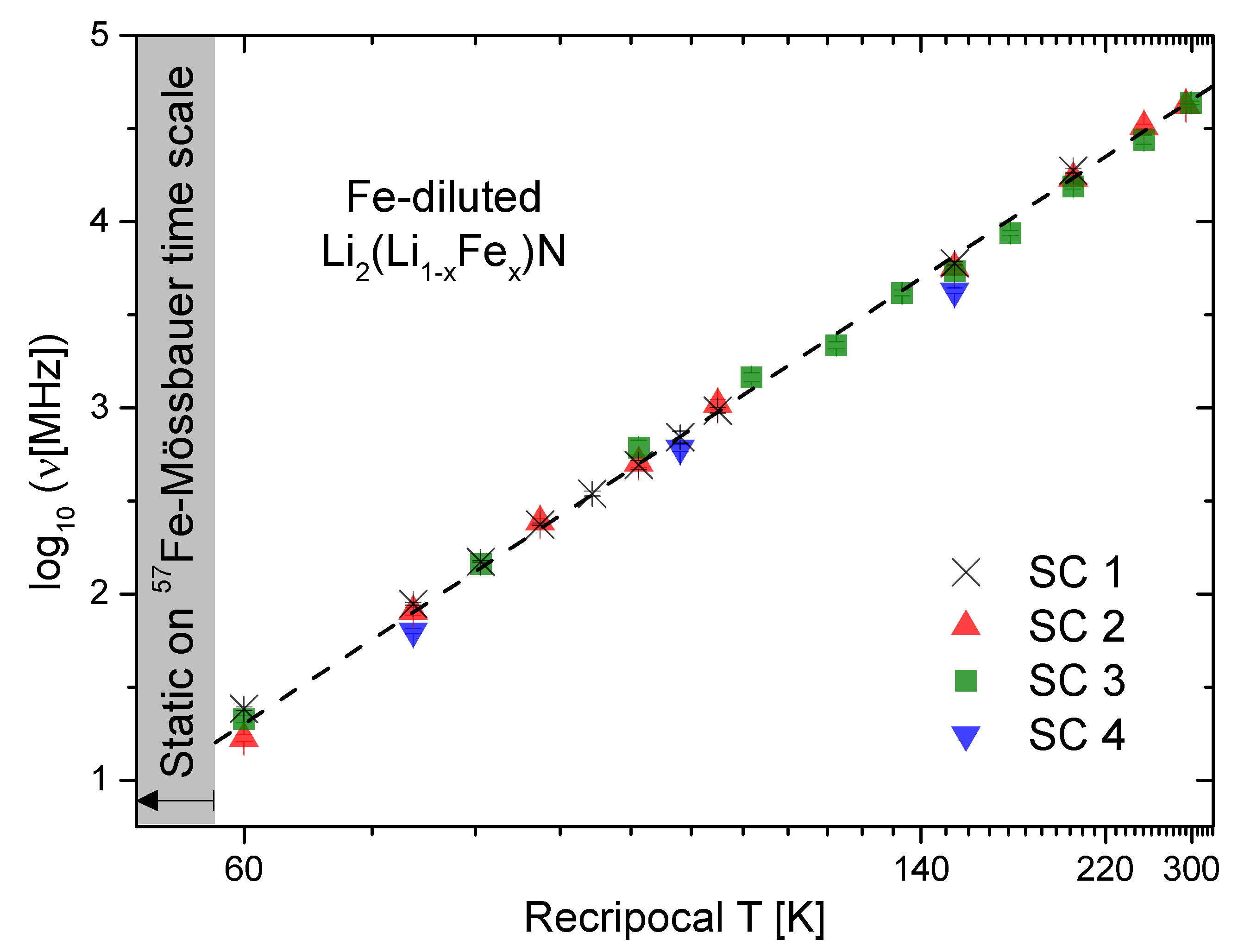}
\caption{\label{fig:row}Logarithmic frequency $\log_{10}\nu$ of Fe site A of SC 1-4 as a function of temperature.}
\end{figure}

Fig.~\ref{fig:row} shows the Arrhenius plot (reciprocal $T$-scaling)
\begin{equation}
\ln\nu=\ln\nu_0-\frac{E_A}{k_B}\left(\frac{1}{T}\right)
\end{equation}
of the extracted fluctuation frequencies $\ln\nu$ of SC 1-4 in MHz. 
The fluctuation frequency is concentration independent as reflected by the parameter $E_B$ and $\ln\nu_0$ of table \ref{table6}.

\subsection{Land\'e factor}
To estimate the Zeeman splitting, it is important to recall the large effective magnetic moment $\mu_\mathrm{eff}=6.5(4)\mu_B$ per Fe atom parallel to the $c$-axis  \cite{Jesche2014b} which is close to the full Hunds' rule value of Fe$^+$. This indicates the validity of the Hunds' rules in this system. Using Russel-Saunders coupling and the proposed spin quantum number $L=2$ and $S=3/2$, we get the Land\'e factor for $J=7/2$,
\begin{equation}
g_J=\frac{3}{2}+\frac{S(S+1)-L(L+1)}{2J(J+1)}\approx1.55.
\end{equation}
Here, $g_L=1$ and $g_S=2$ are used. 

\subsection{Comparison with ferrous halides}
\begin{figure}
\includegraphics[width=\columnwidth]{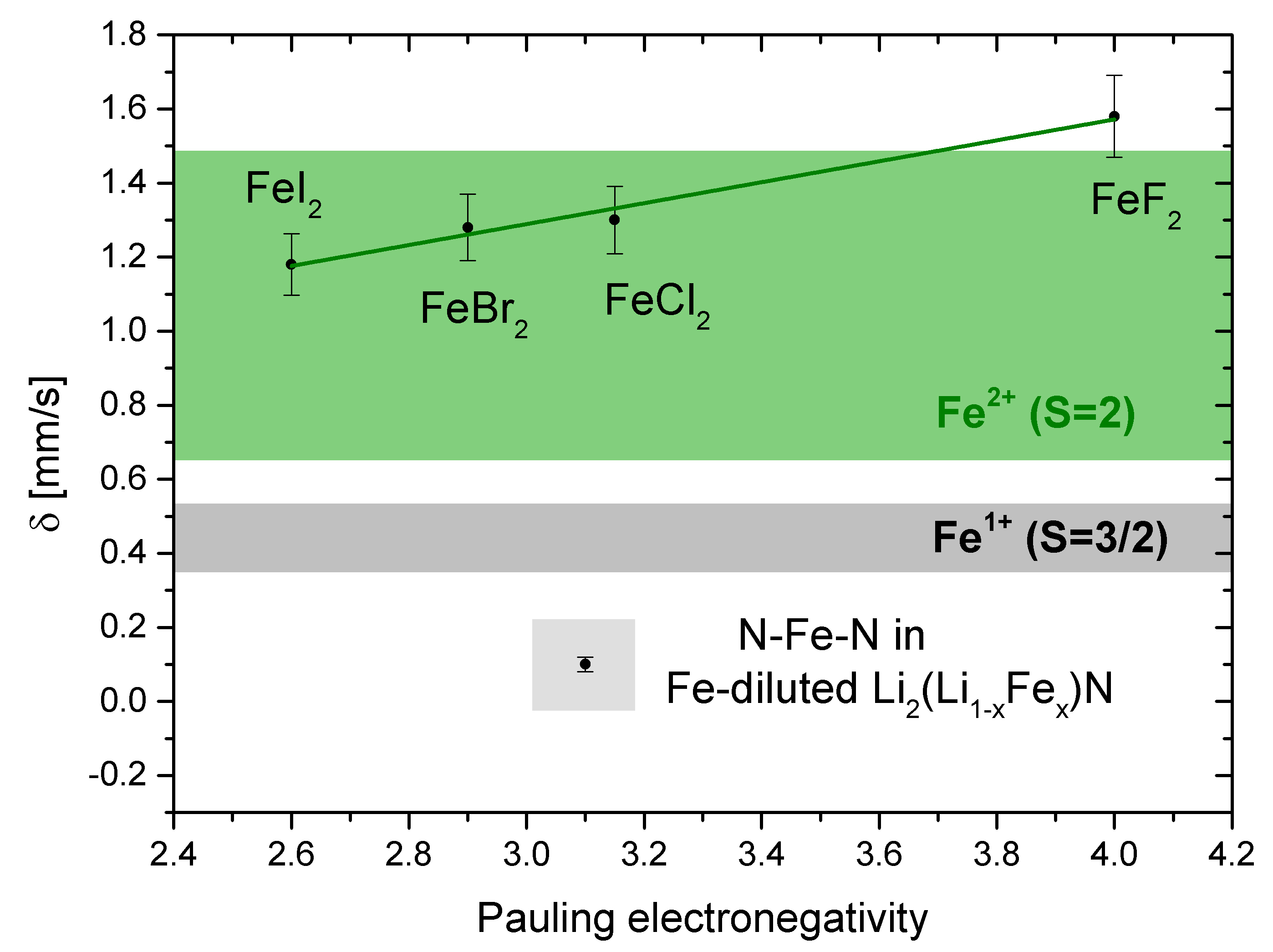}
\caption{\label{fig:halide}Isomer shift relative to $\alpha$-iron as a function of the Pauling electronegativity of the ferrous halides compared with SC 1 
\cite{1968CPL.....2..673A}.}
\end{figure}

Axtmann \textit{et al.} have found a linear relationship between the Pauling electronegativity and the isomer shift $\delta$ in ferrous halides is discussed \cite{1968CPL.....2..673A}. The difference of the ligand electronegativity is related to the isomer shift. This is treated as direct evidence for the participation of $4s$ electrons in the formation of the chemical bonds~\cite{chen_mossbauer_2007}. Fig.\,\ref{fig:halide} shows the presented ferrous halides by Axtmann \textit{et al.} and the isomer shift of SC 1. The observed isomer shift deviates strongly. In the ferrous halides the electronic configuration is $3d4s^x$ where $x$ measures the ionicity. The ionicity increased with $x$~\cite{chen_mossbauer_2007}. The electric monopole hyperfine interaction in \Li is far away from the values of the Fe$^{2+}$ ferrous halides. The quadrupole splitting in the ferrous halides behaves linear as a function of the isomer shift as well~\cite{1968CPL.....2..673A}. The values are between 1.4\,mm/s (FeI$_2$) and 2.6\,mm/s (FeF$_2$). For conversion~\cite{2016HyInt.237...82K} one can use
\begin{equation}
\Delta v_{QS}=\frac{ceQ}{2E_\gamma}V_{zz}\sqrt{1+\frac{\eta^2}{3}}
\end{equation}
with
\begin{equation}
\frac{ceQ}{2E_\gamma}\approx 0.0167\frac{\mathrm{mm/s}}{\mathrm{V/\AA^2}}.
\end{equation}
The value of Fe-diluted \Li is -2.572(2)\,mm/s which shows a comparable electric quadrupole hyperfine interaction with respect to the amount of $V_{zz}$.
\begin{figure}
\includegraphics[width=\columnwidth]{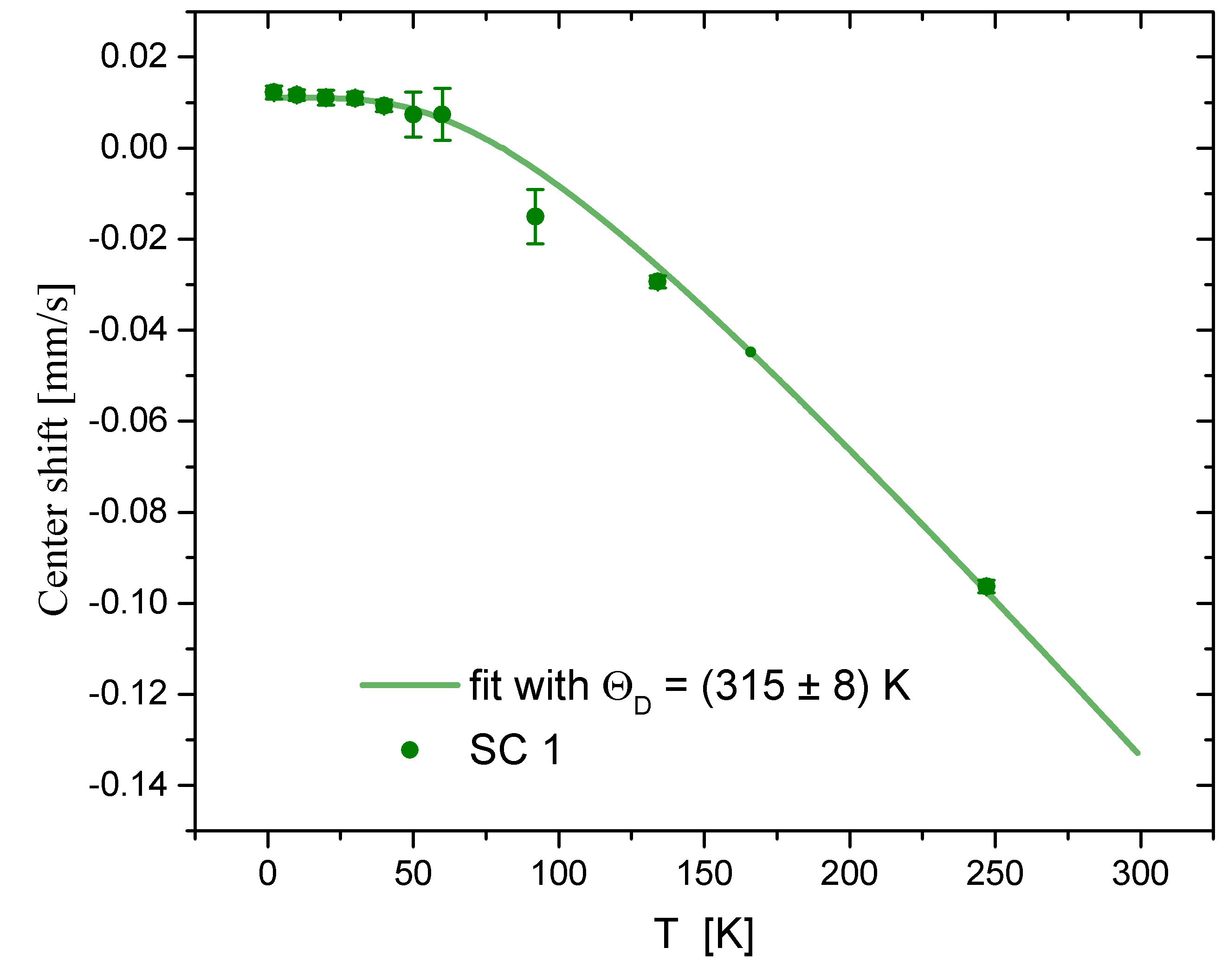}
\caption{\label{fig:debye}Center shift as a function of the temperature and fit to determine the Debye temperature $\Theta_D$.}
\end{figure}

\subsection{Determination of the Debye-temperature $\Theta_D$}

Fig.~\ref{fig:debye} shows the center shift as a function of temperature obtained in ZF of SC 1. The center shift is here $\delta+\delta_\mathrm{SOD}$ without $\alpha$-Iron correction and therefore relative to the $^{57}$Co-source. The temperature dependence of SC 1 yielded a Debye-temperature of $\Theta_D=315(8)$\,K which is a measure of the collective motion of the surrounding atoms of the M\"ossbauer nucleus. One should keep in mind the special geometry with the $\gamma$-beam parallel to the crystallographic c-axis and therefore the phonic excitations in c-direction are considered according to the Debye-Waller factor. In table~\ref{table5} we compare this value with the aforementioned linear C-Fe-C compounds. The values for $[$Fe(C(SiMe$_3$)$_3$)$_2$$]$$^{1-}$ and \Li are similar. This fact further supports the conclusion of a similar electronic configuration of the Fe ion drawn from the values of the quadrupole splitting $\Delta E_Q$ and the magnetic hyperfine field $B_A$ in \Li compared to those of $[$Fe(C(SiMe$_3$)$_3$)$_2$$]$$^{1-}$. 
\begin{table}
\begin{tabular}{c|c}
Compound&$\Theta_D$ [K]\\\hline
$[$Fe(C(SiMe$_3$)$_3$)$_2$]$^{1-}$&313(16)\\
$[$Fe(C(SiMe$_3$)$_3$)$_2$]&125(1)\\
Li$_2$(Li$_{1-x}$Fe$_x$)N&315(8)\\\hline
\end{tabular}
\caption{Comparison of the Debye-temperature $\Theta_D$ for \Li with results for similar linear Fe$^{1+}$ and Fe$^{2+}$ complexes  ~\cite{zadrozny_mossbauer_2013}.}\label{table5}
\end{table}
\subsection{$^{57}$Fe-M\"ossbauer measurement at 80\,K in 2\,T LF of \Li}
\begin{figure}
\includegraphics[width=\columnwidth]{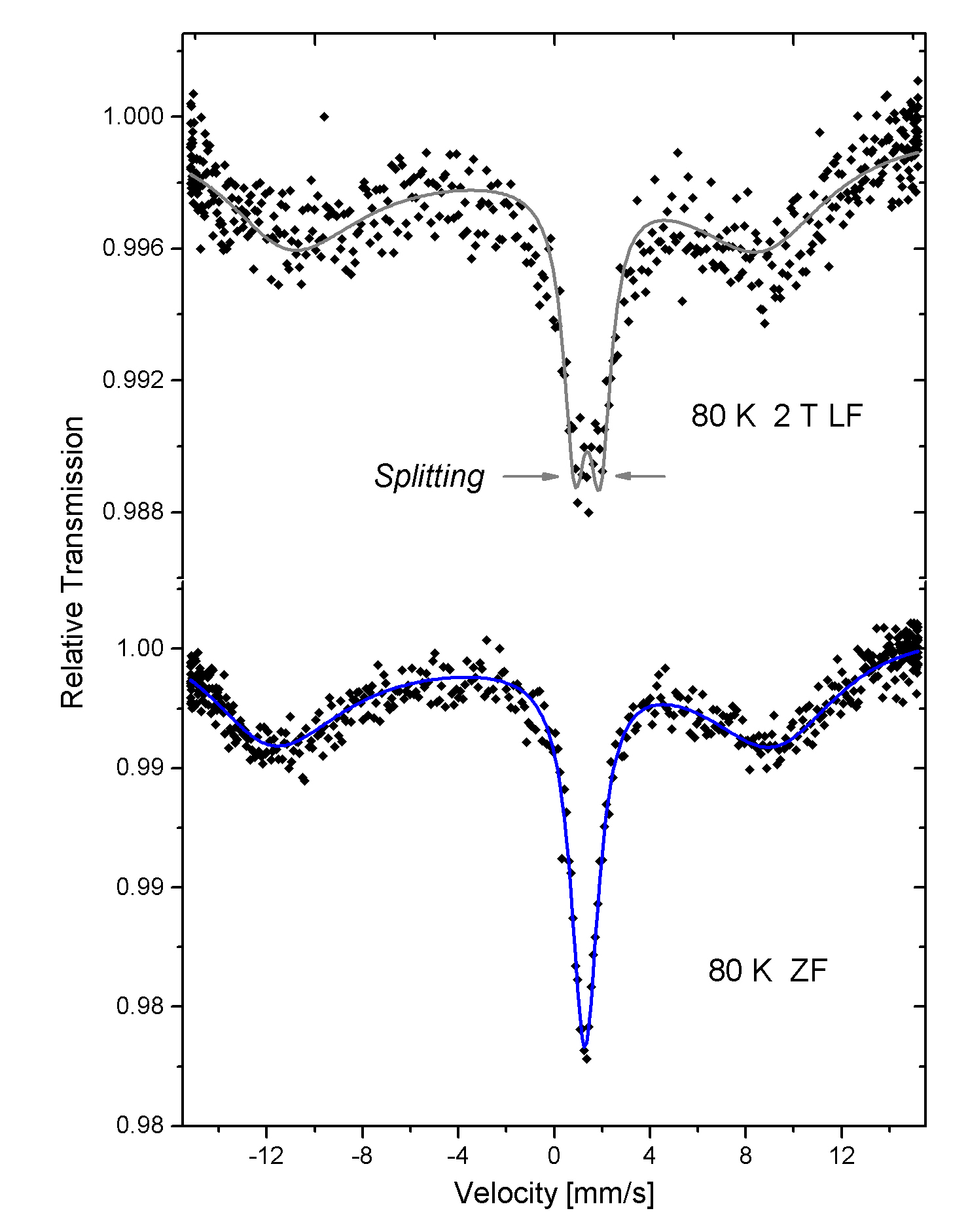}
\caption{\label{fig:80KLF}Comparison of the M\"ossbauer measurements at 80\,K in ZF and in an applied longitudinal field of 2\,T. The fit model is described in the main text.}
\end{figure}

Fig.~\ref{fig:80KLF} shows a comparison of the $^{57}$Fe-M\"ossbauer measurements at 80\,K in ZF and in an applied longitudinal field of 2\,T of SC 1. The analysis is done in the same way as discussed in the main text. The intermediate relaxation line splits  at 80\,K in a magnetic field of 2\,T LF. The doublet is weakly adumbrated because of the scattering of the data (lower magnitude of absorption) and not so well pronounced like in the presented 100\,K measurement at 2\,T. However, a splitting is confirmed. The grey fit is the result of free convergence of the mentioned two-frequency spin reversal model. The relaxation frequencies are $\log_{10}\nu_\mathrm{up}=2.63(8)\log_{10}$[MHz] and $\log_{10}\nu_\mathrm{down}=2.36(6)\log_{10}$[MHz], therefore $\nu_\mathrm{up}\approx427$\,MHz and $\nu_\mathrm{down}\approx229$\,MHz.

\bibliography{Li2LiFe57N}

\begin{thebibliography}{46}%
\makeatletter
\providecommand \@ifxundefined [1]{%
 \@ifx{#1\undefined}
}%
\providecommand \@ifnum [1]{%
 \ifnum #1\expandafter \@firstoftwo
 \else \expandafter \@secondoftwo
 \fi
}%
\providecommand \@ifx [1]{%
 \ifx #1\expandafter \@firstoftwo
 \else \expandafter \@secondoftwo
 \fi
}%
\providecommand \natexlab [1]{#1}%
\providecommand \enquote  [1]{``#1''}%
\providecommand \bibnamefont  [1]{#1}%
\providecommand \bibfnamefont [1]{#1}%
\providecommand \citenamefont [1]{#1}%
\providecommand \href@noop [0]{\@secondoftwo}%
\providecommand \href [0]{\begingroup \@sanitize@url \@href}%
\providecommand \@href[1]{\@@startlink{#1}\@@href}%
\providecommand \@@href[1]{\endgroup#1\@@endlink}%
\providecommand \@sanitize@url [0]{\catcode `\\12\catcode `\$12\catcode
  `\&12\catcode `\#12\catcode `\^12\catcode `\_12\catcode `\%12\relax}%
\providecommand \@@startlink[1]{}%
\providecommand \@@endlink[0]{}%
\providecommand \url  [0]{\begingroup\@sanitize@url \@url }%
\providecommand \@url [1]{\endgroup\@href {#1}{\urlprefix }}%
\providecommand \urlprefix  [0]{URL }%
\providecommand \Eprint [0]{\href }%
\providecommand \doibase [0]{http://dx.doi.org/}%
\providecommand \selectlanguage [0]{\@gobble}%
\providecommand \bibinfo  [0]{\@secondoftwo}%
\providecommand \bibfield  [0]{\@secondoftwo}%
\providecommand \translation [1]{[#1]}%
\providecommand \BibitemOpen [0]{}%
\providecommand \bibitemStop [0]{}%
\providecommand \bibitemNoStop [0]{.\EOS\space}%
\providecommand \EOS [0]{\spacefactor3000\relax}%
\providecommand \BibitemShut  [1]{\csname bibitem#1\endcsname}%
\let\auto@bib@innerbib\@empty
\bibitem [{\citenamefont {Gatteschi}\ and\ \citenamefont
  {Sessoli}(2003)}]{ANIE:ANIE200390099}%
  \BibitemOpen
  \bibfield  {author} {\bibinfo {author} {\bibfnamefont {D.}~\bibnamefont
  {Gatteschi}}\ and\ \bibinfo {author} {\bibfnamefont {R.}~\bibnamefont
  {Sessoli}},\ }\href {\doibase 10.1002/anie.200390099} {\bibfield  {journal}
  {\bibinfo  {journal} {Angewandte Chemie International Edition}\ }\textbf
  {\bibinfo {volume} {42}},\ \bibinfo {pages} {268} (\bibinfo {year}
  {2003})}\BibitemShut {NoStop}%
\bibitem [{\citenamefont {Gatteschi}\ \emph {et~al.}(2006)\citenamefont
  {Gatteschi}, \citenamefont {Sessoli},\ and\ \citenamefont
  {Villain}}]{gatteschi_molecular_2006}%
  \BibitemOpen
  \bibfield  {author} {\bibinfo {author} {\bibfnamefont {D.}~\bibnamefont
  {Gatteschi}}, \bibinfo {author} {\bibfnamefont {R.}~\bibnamefont {Sessoli}},
  \ and\ \bibinfo {author} {\bibfnamefont {J.}~\bibnamefont {Villain}},\ }\href
  {http://www.oxfordscholarship.com/view/10.1093/acprof:oso/9780198567530.001.0001/acprof-9780198567530}
  {\emph {\bibinfo {title} {Molecular Nanomagnets}}}\ (\bibinfo  {publisher}
  {Oxford University Press},\ \bibinfo {year} {2006})\BibitemShut {NoStop}%
\bibitem [{\citenamefont {Bogani}\ and\ \citenamefont
  {Wernsdorfer}(2008)}]{bogani_molecular_2008}%
  \BibitemOpen
  \bibfield  {author} {\bibinfo {author} {\bibfnamefont {L.}~\bibnamefont
  {Bogani}}\ and\ \bibinfo {author} {\bibfnamefont {W.}~\bibnamefont
  {Wernsdorfer}},\ }\href {\doibase 10.1038/nmat2133} {\bibfield  {journal}
  {\bibinfo  {journal} {Nature Materials}\ }\textbf {\bibinfo {volume} {7}},\
  \bibinfo {pages} {179} (\bibinfo {year} {2008})}\BibitemShut {NoStop}%
\bibitem [{\citenamefont {Strandberg}\ \emph {et~al.}(2007)\citenamefont
  {Strandberg}, \citenamefont {Canali},\ and\ \citenamefont
  {{MacDonald}}}]{strandberg_transition-metal_2007}%
  \BibitemOpen
  \bibfield  {author} {\bibinfo {author} {\bibfnamefont {T.~O.}\ \bibnamefont
  {Strandberg}}, \bibinfo {author} {\bibfnamefont {C.~M.}\ \bibnamefont
  {Canali}}, \ and\ \bibinfo {author} {\bibfnamefont {A.~H.}\ \bibnamefont
  {{MacDonald}}},\ }\href {\doibase 10.1038/nmat1968} {\bibfield  {journal}
  {\bibinfo  {journal} {Nature Materials}\ }\textbf {\bibinfo {volume} {6}},\
  \bibinfo {pages} {648} (\bibinfo {year} {2007})}\BibitemShut {NoStop}%
\bibitem [{\citenamefont {Leuenberger}\ and\ \citenamefont
  {Loss}(2001)}]{leuenberger_quantum_2001}%
  \BibitemOpen
  \bibfield  {author} {\bibinfo {author} {\bibfnamefont {M.~N.}\ \bibnamefont
  {Leuenberger}}\ and\ \bibinfo {author} {\bibfnamefont {D.}~\bibnamefont
  {Loss}},\ }\href {\doibase 10.1038/35071024} {\bibfield  {journal} {\bibinfo
  {journal} {Nature}\ }\textbf {\bibinfo {volume} {410}},\ \bibinfo {pages}
  {789} (\bibinfo {year} {2001})}\BibitemShut {NoStop}%
\bibitem [{\citenamefont {Zener}(1932)}]{Zener1932}%
  \BibitemOpen
  \bibfield  {author} {\bibinfo {author} {\bibfnamefont {C.}~\bibnamefont
  {Zener}},\ }\href {\doibase 10.1098/rspa.1932.0165} {\bibfield  {journal}
  {\bibinfo  {journal} {Proc. R. Soc. A}\ }\textbf {\bibinfo {volume} {137}},\
  \bibinfo {pages} {696} (\bibinfo {year} {1932})}\BibitemShut {NoStop}%
\bibitem [{\citenamefont {Thomas}\ \emph {et~al.}(1996)\citenamefont {Thomas},
  \citenamefont {Lionti}, \citenamefont {Ballou}, \citenamefont {Gatteschi},
  \citenamefont {Sessoli},\ and\ \citenamefont
  {Barbara}}]{thomas_macroscopic_1996}%
  \BibitemOpen
  \bibfield  {author} {\bibinfo {author} {\bibfnamefont {L.}~\bibnamefont
  {Thomas}}, \bibinfo {author} {\bibfnamefont {F.}~\bibnamefont {Lionti}},
  \bibinfo {author} {\bibfnamefont {R.}~\bibnamefont {Ballou}}, \bibinfo
  {author} {\bibfnamefont {D.}~\bibnamefont {Gatteschi}}, \bibinfo {author}
  {\bibfnamefont {R.}~\bibnamefont {Sessoli}}, \ and\ \bibinfo {author}
  {\bibfnamefont {B.}~\bibnamefont {Barbara}},\ }\href {\doibase
  10.1038/383145a0} {\bibfield  {journal} {\bibinfo  {journal} {Nature}\
  }\textbf {\bibinfo {volume} {383}},\ \bibinfo {pages} {145} (\bibinfo {year}
  {1996})}\BibitemShut {NoStop}%
\bibitem [{\citenamefont {Sangregorio}\ \emph {et~al.}(1997)\citenamefont
  {Sangregorio}, \citenamefont {Ohm}, \citenamefont {Paulsen}, \citenamefont
  {Sessoli},\ and\ \citenamefont {Gatteschi}}]{Sangregorio1997}%
  \BibitemOpen
  \bibfield  {author} {\bibinfo {author} {\bibfnamefont {C.}~\bibnamefont
  {Sangregorio}}, \bibinfo {author} {\bibfnamefont {T.}~\bibnamefont {Ohm}},
  \bibinfo {author} {\bibfnamefont {C.}~\bibnamefont {Paulsen}}, \bibinfo
  {author} {\bibfnamefont {R.}~\bibnamefont {Sessoli}}, \ and\ \bibinfo
  {author} {\bibfnamefont {D.}~\bibnamefont {Gatteschi}},\ }\href {\doibase
  10.1103/PhysRevLett.78.4645} {\bibfield  {journal} {\bibinfo  {journal}
  {Phys. Rev. Lett.}\ }\textbf {\bibinfo {volume} {78}},\ \bibinfo {pages}
  {4645} (\bibinfo {year} {1997})}\BibitemShut {NoStop}%
\bibitem [{\citenamefont {Loss}\ \emph {et~al.}(1992)\citenamefont {Loss},
  \citenamefont {DiVincenzo},\ and\ \citenamefont
  {Grinstein}}]{PhysRevLett.69.3232}%
  \BibitemOpen
  \bibfield  {author} {\bibinfo {author} {\bibfnamefont {D.}~\bibnamefont
  {Loss}}, \bibinfo {author} {\bibfnamefont {D.~P.}\ \bibnamefont
  {DiVincenzo}}, \ and\ \bibinfo {author} {\bibfnamefont {G.}~\bibnamefont
  {Grinstein}},\ }\href {\doibase 10.1103/PhysRevLett.69.3232} {\bibfield
  {journal} {\bibinfo  {journal} {Phys. Rev. Lett.}\ }\textbf {\bibinfo
  {volume} {69}},\ \bibinfo {pages} {3232} (\bibinfo {year}
  {1992})}\BibitemShut {NoStop}%
\bibitem [{\citenamefont {Wernsdorfer}\ and\ \citenamefont
  {Sessoli}(1999)}]{Wernsdorfer1999}%
  \BibitemOpen
  \bibfield  {author} {\bibinfo {author} {\bibfnamefont {W.}~\bibnamefont
  {Wernsdorfer}}\ and\ \bibinfo {author} {\bibfnamefont {R.}~\bibnamefont
  {Sessoli}},\ }\href {\doibase 10.1126/science.284.5411.133} {\bibfield
  {journal} {\bibinfo  {journal} {Science}\ }\textbf {\bibinfo {volume}
  {284}},\ \bibinfo {pages} {133} (\bibinfo {year} {1999})}\BibitemShut
  {NoStop}%
\bibitem [{\citenamefont {Klatyk}\ and\ \citenamefont
  {Kniep}(1999)}]{Klatyk1999}%
  \BibitemOpen
  \bibfield  {author} {\bibinfo {author} {\bibfnamefont {J.}~\bibnamefont
  {Klatyk}}\ and\ \bibinfo {author} {\bibfnamefont {R.}~\bibnamefont {Kniep}},\
  }\href@noop {} {\bibfield  {journal} {\bibinfo  {journal} {Z. Kristallogr. -
  New Cryst. Struct.}\ }\textbf {\bibinfo {volume} {214}},\ \bibinfo {pages}
  {447} (\bibinfo {year} {1999})}\BibitemShut {NoStop}%
\bibitem [{\citenamefont {Klatyk}\ \emph
  {et~al.}(2002{\natexlab{a}})\citenamefont {Klatyk}, \citenamefont {Schnelle},
  \citenamefont {Wagner}, \citenamefont {Niewa}, \citenamefont {Nov\'ak},
  \citenamefont {Kniep}, \citenamefont {Waldeck}, \citenamefont {Ksenofontov},\
  and\ \citenamefont {G\"utlich}}]{PhysRevLett.88.207202}%
  \BibitemOpen
  \bibfield  {author} {\bibinfo {author} {\bibfnamefont {J.}~\bibnamefont
  {Klatyk}}, \bibinfo {author} {\bibfnamefont {W.}~\bibnamefont {Schnelle}},
  \bibinfo {author} {\bibfnamefont {F.~R.}\ \bibnamefont {Wagner}}, \bibinfo
  {author} {\bibfnamefont {R.}~\bibnamefont {Niewa}}, \bibinfo {author}
  {\bibfnamefont {P.}~\bibnamefont {Nov\'ak}}, \bibinfo {author} {\bibfnamefont
  {R.}~\bibnamefont {Kniep}}, \bibinfo {author} {\bibfnamefont
  {M.}~\bibnamefont {Waldeck}}, \bibinfo {author} {\bibfnamefont
  {V.}~\bibnamefont {Ksenofontov}}, \ and\ \bibinfo {author} {\bibfnamefont
  {P.}~\bibnamefont {G\"utlich}},\ }\href {\doibase
  10.1103/PhysRevLett.88.207202} {\bibfield  {journal} {\bibinfo  {journal}
  {Phys. Rev. Lett.}\ }\textbf {\bibinfo {volume} {88}},\ \bibinfo {pages}
  {207202} (\bibinfo {year} {2002}{\natexlab{a}})}\BibitemShut {NoStop}%
\bibitem [{\citenamefont {Ksenofontov}\ \emph {et~al.}(2003)\citenamefont
  {Ksenofontov}, \citenamefont {Reiman}, \citenamefont {Waldeck}, \citenamefont
  {Niewa}, \citenamefont {Kniep},\ and\ \citenamefont
  {G\"utlich}}]{ZAAC:ZAAC200300135}%
  \BibitemOpen
  \bibfield  {author} {\bibinfo {author} {\bibfnamefont {V.}~\bibnamefont
  {Ksenofontov}}, \bibinfo {author} {\bibfnamefont {S.}~\bibnamefont {Reiman}},
  \bibinfo {author} {\bibfnamefont {M.}~\bibnamefont {Waldeck}}, \bibinfo
  {author} {\bibfnamefont {R.}~\bibnamefont {Niewa}}, \bibinfo {author}
  {\bibfnamefont {R.}~\bibnamefont {Kniep}}, \ and\ \bibinfo {author}
  {\bibfnamefont {P.}~\bibnamefont {G\"utlich}},\ }\href {\doibase
  10.1002/zaac.200300135} {\bibfield  {journal} {\bibinfo  {journal}
  {Zeitschrift f\"ur anorganische und allgemeine Chemie}\ }\textbf {\bibinfo
  {volume} {629}},\ \bibinfo {pages} {1787} (\bibinfo {year}
  {2003})}\BibitemShut {NoStop}%
\bibitem [{\citenamefont {Jesche}\ \emph {et~al.}(2014)\citenamefont {Jesche},
  \citenamefont {McCallum}, \citenamefont {Thimmaiah}, \citenamefont {Jacobs},
  \citenamefont {Taufour}, \citenamefont {Kreyssig}, \citenamefont {Houk},
  \citenamefont {Bud'ko},\ and\ \citenamefont {Canfield}}]{Jesche2014b}%
  \BibitemOpen
  \bibfield  {author} {\bibinfo {author} {\bibfnamefont {A.}~\bibnamefont
  {Jesche}}, \bibinfo {author} {\bibfnamefont {R.~W.}\ \bibnamefont
  {McCallum}}, \bibinfo {author} {\bibfnamefont {S.}~\bibnamefont {Thimmaiah}},
  \bibinfo {author} {\bibfnamefont {J.~L.}\ \bibnamefont {Jacobs}}, \bibinfo
  {author} {\bibfnamefont {V.}~\bibnamefont {Taufour}}, \bibinfo {author}
  {\bibfnamefont {A.}~\bibnamefont {Kreyssig}}, \bibinfo {author}
  {\bibfnamefont {R.~S.}\ \bibnamefont {Houk}}, \bibinfo {author}
  {\bibfnamefont {S.~L.}\ \bibnamefont {Bud'ko}}, \ and\ \bibinfo {author}
  {\bibfnamefont {P.~C.}\ \bibnamefont {Canfield}},\ }\href
  {http://dx.doi.org/10.1038/ncomms4333} {\bibfield  {journal} {\bibinfo
  {journal} {Nat. Commun.}\ }\textbf {\bibinfo {volume} {5:3333}} (\bibinfo
  {year} {2014})},\ \bibinfo {note} {doi: 10.1038/ncomms4333}\BibitemShut
  {NoStop}%
\bibitem [{\citenamefont {Fix}\ \emph {et~al.}(2018{\natexlab{a}})\citenamefont
  {Fix}, \citenamefont {Jesche}, \citenamefont {Jantz}, \citenamefont
  {Br\"auninger}, \citenamefont {Klauss}, \citenamefont {Manna}, \citenamefont
  {Pietsch}, \citenamefont {H\"oppe},\ and\ \citenamefont
  {Canfield}}]{Fix2018b}%
  \BibitemOpen
  \bibfield  {author} {\bibinfo {author} {\bibfnamefont {M.}~\bibnamefont
  {Fix}}, \bibinfo {author} {\bibfnamefont {A.}~\bibnamefont {Jesche}},
  \bibinfo {author} {\bibfnamefont {S.~G.}\ \bibnamefont {Jantz}}, \bibinfo
  {author} {\bibfnamefont {S.~A.}\ \bibnamefont {Br\"auninger}}, \bibinfo
  {author} {\bibfnamefont {H.-H.}\ \bibnamefont {Klauss}}, \bibinfo {author}
  {\bibfnamefont {R.~S.}\ \bibnamefont {Manna}}, \bibinfo {author}
  {\bibfnamefont {I.~M.}\ \bibnamefont {Pietsch}}, \bibinfo {author}
  {\bibfnamefont {H.~A.}\ \bibnamefont {H\"oppe}}, \ and\ \bibinfo {author}
  {\bibfnamefont {P.~C.}\ \bibnamefont {Canfield}},\ }\href {\doibase
  10.1103/PhysRevB.97.064419} {\bibfield  {journal} {\bibinfo  {journal} {Phys.
  Rev. B}\ }\textbf {\bibinfo {volume} {97}},\ \bibinfo {pages} {064419}
  (\bibinfo {year} {2018}{\natexlab{a}})}\BibitemShut {NoStop}%
\bibitem [{\citenamefont {Fix}\ \emph {et~al.}(2018{\natexlab{b}})\citenamefont
  {Fix}, \citenamefont {Atkinson}, \citenamefont {Canfield}, \citenamefont {del
  Barco},\ and\ \citenamefont {Jesche}}]{Fix2018c}%
  \BibitemOpen
  \bibfield  {author} {\bibinfo {author} {\bibfnamefont {M.}~\bibnamefont
  {Fix}}, \bibinfo {author} {\bibfnamefont {J.~H.}\ \bibnamefont {Atkinson}},
  \bibinfo {author} {\bibfnamefont {P.~C.}\ \bibnamefont {Canfield}}, \bibinfo
  {author} {\bibfnamefont {E.}~\bibnamefont {del Barco}}, \ and\ \bibinfo
  {author} {\bibfnamefont {A.}~\bibnamefont {Jesche}},\ }\href {\doibase
  10.1103/PhysRevLett.120.147202} {\bibfield  {journal} {\bibinfo  {journal}
  {Phys. Rev. Lett.}\ }\textbf {\bibinfo {volume} {120}},\ \bibinfo {pages}
  {147202} (\bibinfo {year} {2018}{\natexlab{b}})}\BibitemShut {NoStop}%
\bibitem [{\citenamefont {Klatyk}\ \emph
  {et~al.}(2002{\natexlab{b}})\citenamefont {Klatyk}, \citenamefont {Schnelle},
  \citenamefont {Wagner}, \citenamefont {Niewa}, \citenamefont {Nov\'ak},
  \citenamefont {Kniep}, \citenamefont {Waldeck}, \citenamefont {Ksenofontov},\
  and\ \citenamefont {G\"utlich}}]{Klatyk2002}%
  \BibitemOpen
  \bibfield  {author} {\bibinfo {author} {\bibfnamefont {J.}~\bibnamefont
  {Klatyk}}, \bibinfo {author} {\bibfnamefont {W.}~\bibnamefont {Schnelle}},
  \bibinfo {author} {\bibfnamefont {F.~R.}\ \bibnamefont {Wagner}}, \bibinfo
  {author} {\bibfnamefont {R.}~\bibnamefont {Niewa}}, \bibinfo {author}
  {\bibfnamefont {P.}~\bibnamefont {Nov\'ak}}, \bibinfo {author} {\bibfnamefont
  {R.}~\bibnamefont {Kniep}}, \bibinfo {author} {\bibfnamefont
  {M.}~\bibnamefont {Waldeck}}, \bibinfo {author} {\bibfnamefont
  {V.}~\bibnamefont {Ksenofontov}}, \ and\ \bibinfo {author} {\bibfnamefont
  {P.}~\bibnamefont {G\"utlich}},\ }\href@noop {} {\bibfield  {journal}
  {\bibinfo  {journal} {Phys. Rev. Lett.}\ }\textbf {\bibinfo {volume} {88}},\
  \bibinfo {pages} {207202} (\bibinfo {year} {2002}{\natexlab{b}})}\BibitemShut
  {NoStop}%
\bibitem [{\citenamefont {Xu}\ \emph {et~al.}(2017)\citenamefont {Xu},
  \citenamefont {Zangeneh}, \citenamefont {Yadav}, \citenamefont {Avdoshenko},
  \citenamefont {van~den Brink}, \citenamefont {Jesche},\ and\ \citenamefont
  {Hozoi}}]{C7NR03041J}%
  \BibitemOpen
  \bibfield  {author} {\bibinfo {author} {\bibfnamefont {L.}~\bibnamefont
  {Xu}}, \bibinfo {author} {\bibfnamefont {Z.}~\bibnamefont {Zangeneh}},
  \bibinfo {author} {\bibfnamefont {R.}~\bibnamefont {Yadav}}, \bibinfo
  {author} {\bibfnamefont {S.}~\bibnamefont {Avdoshenko}}, \bibinfo {author}
  {\bibfnamefont {J.}~\bibnamefont {van~den Brink}}, \bibinfo {author}
  {\bibfnamefont {A.}~\bibnamefont {Jesche}}, \ and\ \bibinfo {author}
  {\bibfnamefont {L.}~\bibnamefont {Hozoi}},\ }\href {\doibase
  10.1039/C7NR03041J} {\bibfield  {journal} {\bibinfo  {journal} {Nanoscale}\
  }\textbf {\bibinfo {volume} {9}},\ \bibinfo {pages} {10596} (\bibinfo {year}
  {2017})}\BibitemShut {NoStop}%
\bibitem [{\citenamefont {Jesche}\ and\ \citenamefont
  {Canfield}(2014)}]{Jesche2014c}%
  \BibitemOpen
  \bibfield  {author} {\bibinfo {author} {\bibfnamefont {A.}~\bibnamefont
  {Jesche}}\ and\ \bibinfo {author} {\bibfnamefont {P.~C.}\ \bibnamefont
  {Canfield}},\ }\href {\doibase 10.1080/14786435.2014.913114} {\bibfield
  {journal} {\bibinfo  {journal} {Philos. Mag.}\ }\textbf {\bibinfo {volume}
  {94}},\ \bibinfo {pages} {2372} (\bibinfo {year} {2014})}\BibitemShut
  {NoStop}%
\bibitem [{\citenamefont {Canfield}\ and\ \citenamefont
  {Fisher}(2001)}]{Canfield2001}%
  \BibitemOpen
  \bibfield  {author} {\bibinfo {author} {\bibfnamefont {P.~C.}\ \bibnamefont
  {Canfield}}\ and\ \bibinfo {author} {\bibfnamefont {I.~R.}\ \bibnamefont
  {Fisher}},\ }\href {\doibase 10.1016/S0022-0248(01)00827-2} {\bibfield
  {journal} {\bibinfo  {journal} {J. Cryst. Growth}\ }\textbf {\bibinfo
  {volume} {225}},\ \bibinfo {pages} {155} (\bibinfo {year}
  {2001})}\BibitemShut {NoStop}%
\bibitem [{\citenamefont {{Kamusella}}\ and\ \citenamefont
  {{Klauss}}(2016)}]{2016HyInt.237...82K}%
  \BibitemOpen
  \bibfield  {author} {\bibinfo {author} {\bibfnamefont {S.}~\bibnamefont
  {{Kamusella}}}\ and\ \bibinfo {author} {\bibfnamefont {H.-H.}\ \bibnamefont
  {{Klauss}}},\ }\href {\doibase 10.1007/s10751-016-1247-z} {\bibfield
  {journal} {\bibinfo  {journal} {Hyperfine Interactions}\ }\textbf {\bibinfo
  {volume} {237}},\ \bibinfo {eid} {82} (\bibinfo {year} {2016})}\BibitemShut
  {NoStop}%
\bibitem [{\citenamefont {Chen}\ and\ \citenamefont
  {Yang}(2007)}]{chen_mossbauer_2007}%
  \BibitemOpen
  \bibfield  {author} {\bibinfo {author} {\bibfnamefont {Y.}~\bibnamefont
  {Chen}}\ and\ \bibinfo {author} {\bibfnamefont {D.-P.}\ \bibnamefont
  {Yang}},\ }\href@noop {} {\emph {\bibinfo {title} {M\"ossbauer effect in
  lattice dynamics: experimental techniques and applications}}}\ (\bibinfo
  {publisher} {Wiley-{VCH} ; John Wiley},\ \bibinfo {year} {2007})\BibitemShut
  {NoStop}%
\bibitem [{\citenamefont {Guetlich}\ \emph {et~al.}(2011)\citenamefont
  {Guetlich}, \citenamefont {Bill},\ and\ \citenamefont
  {Trautwein}}]{gutlich_mossbauer_2011}%
  \BibitemOpen
  \bibfield  {author} {\bibinfo {author} {\bibfnamefont {P.}~\bibnamefont
  {Guetlich}}, \bibinfo {author} {\bibfnamefont {E.}~\bibnamefont {Bill}}, \
  and\ \bibinfo {author} {\bibfnamefont {A.~X.}\ \bibnamefont {Trautwein}},\
  }\href {http://link.springer.com/10.1007/978-3-540-88428-6} {\emph {\bibinfo
  {title} {Moessbauer Spectroscopy and Transition Metal Chemistry}}}\ (\bibinfo
   {publisher} {Springer Berlin Heidelberg},\ \bibinfo {year} {2011})\ \bibinfo
  {note} {{DOI}: 10.1007/978-3-540-88428-6}\BibitemShut {NoStop}%
\bibitem [{\citenamefont {Chuev}(2011)}]{0953-8984-23-42-426003}%
  \BibitemOpen
  \bibfield  {author} {\bibinfo {author} {\bibfnamefont {M.~A.}\ \bibnamefont
  {Chuev}},\ }\href {http://stacks.iop.org/0953-8984/23/i=42/a=426003}
  {\bibfield  {journal} {\bibinfo  {journal} {Journal of Physics: Condensed
  Matter}\ }\textbf {\bibinfo {volume} {23}},\ \bibinfo {pages} {426003}
  (\bibinfo {year} {2011})}\BibitemShut {NoStop}%
\bibitem [{\citenamefont {Faid}\ and\ \citenamefont
  {Fox}(1986)}]{PhysRevA.34.4286}%
  \BibitemOpen
  \bibfield  {author} {\bibinfo {author} {\bibfnamefont {K.}~\bibnamefont
  {Faid}}\ and\ \bibinfo {author} {\bibfnamefont {R.~F.}\ \bibnamefont {Fox}},\
  }\href {\doibase 10.1103/PhysRevA.34.4286} {\bibfield  {journal} {\bibinfo
  {journal} {Phys. Rev. A}\ }\textbf {\bibinfo {volume} {34}},\ \bibinfo
  {pages} {4286} (\bibinfo {year} {1986})}\BibitemShut {NoStop}%
\bibitem [{\citenamefont {Blume}(1968)}]{PhysRev.174.351}%
  \BibitemOpen
  \bibfield  {author} {\bibinfo {author} {\bibfnamefont {M.}~\bibnamefont
  {Blume}},\ }\href {\doibase 10.1103/PhysRev.174.351} {\bibfield  {journal}
  {\bibinfo  {journal} {Phys. Rev.}\ }\textbf {\bibinfo {volume} {174}},\
  \bibinfo {pages} {351} (\bibinfo {year} {1968})}\BibitemShut {NoStop}%
\bibitem [{Car(2006)}]{Carretta2006}%
  \BibitemOpen
  \href@noop {} {\enquote {\bibinfo {title} {{NMR, muSR and M\"ossbauer
  spectroscopy in molecular magnets}},}\ } (\bibinfo {year} {2006}),\ \bibinfo
  {note} {{OCLC}: 255432892}\BibitemShut {NoStop}%
\bibitem [{\citenamefont {Segal}\ and\ \citenamefont
  {Wallace}(1970)}]{SEGAL1970347}%
  \BibitemOpen
  \bibfield  {author} {\bibinfo {author} {\bibfnamefont {E.}~\bibnamefont
  {Segal}}\ and\ \bibinfo {author} {\bibfnamefont {W.}~\bibnamefont
  {Wallace}},\ }\href {\doibase https://doi.org/10.1016/0022-4596(70)90093-9}
  {\bibfield  {journal} {\bibinfo  {journal} {Journal of Solid State
  Chemistry}\ }\textbf {\bibinfo {volume} {2}},\ \bibinfo {pages} {347 }
  (\bibinfo {year} {1970})}\BibitemShut {NoStop}%
\bibitem [{\citenamefont {Segal}\ and\ \citenamefont
  {Wallace}(1973)}]{SEGAL197399}%
  \BibitemOpen
  \bibfield  {author} {\bibinfo {author} {\bibfnamefont {E.}~\bibnamefont
  {Segal}}\ and\ \bibinfo {author} {\bibfnamefont {W.}~\bibnamefont
  {Wallace}},\ }\href {\doibase https://doi.org/10.1016/0022-4596(73)90210-7}
  {\bibfield  {journal} {\bibinfo  {journal} {Journal of Solid State
  Chemistry}\ }\textbf {\bibinfo {volume} {6}},\ \bibinfo {pages} {99 }
  (\bibinfo {year} {1973})}\BibitemShut {NoStop}%
\bibitem [{\citenamefont {Segal}\ and\ \citenamefont
  {Wallace}(1974)}]{SEGAL1974203}%
  \BibitemOpen
  \bibfield  {author} {\bibinfo {author} {\bibfnamefont {E.}~\bibnamefont
  {Segal}}\ and\ \bibinfo {author} {\bibfnamefont {W.}~\bibnamefont
  {Wallace}},\ }\href {\doibase https://doi.org/10.1016/S0022-4596(74)80005-8}
  {\bibfield  {journal} {\bibinfo  {journal} {Journal of Solid State
  Chemistry}\ }\textbf {\bibinfo {volume} {11}},\ \bibinfo {pages} {203 }
  (\bibinfo {year} {1974})}\BibitemShut {NoStop}%
\bibitem [{\citenamefont {Nov\'ak}\ and\ \citenamefont
  {Wagner}(2002)}]{PhysRevB.66.184434}%
  \BibitemOpen
  \bibfield  {author} {\bibinfo {author} {\bibfnamefont {P.}~\bibnamefont
  {Nov\'ak}}\ and\ \bibinfo {author} {\bibfnamefont {F.~R.}\ \bibnamefont
  {Wagner}},\ }\href {\doibase 10.1103/PhysRevB.66.184434} {\bibfield
  {journal} {\bibinfo  {journal} {Phys. Rev. B}\ }\textbf {\bibinfo {volume}
  {66}},\ \bibinfo {pages} {184434} (\bibinfo {year} {2002})}\BibitemShut
  {NoStop}%
\bibitem [{\citenamefont {Ulyanov}\ and\ \citenamefont
  {Zaslavskii}(1992)}]{ULYANOV1992179}%
  \BibitemOpen
  \bibfield  {author} {\bibinfo {author} {\bibfnamefont {V.}~\bibnamefont
  {Ulyanov}}\ and\ \bibinfo {author} {\bibfnamefont {O.}~\bibnamefont
  {Zaslavskii}},\ }\href {\doibase
  https://doi.org/10.1016/0370-1573(92)90158-V} {\bibfield  {journal} {\bibinfo
   {journal} {Physics Reports}\ }\textbf {\bibinfo {volume} {216}},\ \bibinfo
  {pages} {179 } (\bibinfo {year} {1992})}\BibitemShut {NoStop}%
\bibitem [{\citenamefont {Zaslavskii}(1990)}]{Zaslavskii90}%
  \BibitemOpen
  \bibfield  {author} {\bibinfo {author} {\bibfnamefont {O.~B.}\ \bibnamefont
  {Zaslavskii}},\ }\href {\doibase 10.1103/PhysRevB.42.992} {\bibfield
  {journal} {\bibinfo  {journal} {Phys. Rev. B}\ }\textbf {\bibinfo {volume}
  {42}},\ \bibinfo {pages} {992} (\bibinfo {year} {1990})}\BibitemShut
  {NoStop}%
\bibitem [{\citenamefont {Enz}\ and\ \citenamefont
  {Schilling}(1986)}]{0022-3719-19-30-005}%
  \BibitemOpen
  \bibfield  {author} {\bibinfo {author} {\bibfnamefont {M.}~\bibnamefont
  {Enz}}\ and\ \bibinfo {author} {\bibfnamefont {R.}~\bibnamefont
  {Schilling}},\ }\href {http://stacks.iop.org/0022-3719/19/i=30/a=005}
  {\bibfield  {journal} {\bibinfo  {journal} {Journal of Physics C: Solid State
  Physics}\ }\textbf {\bibinfo {volume} {19}},\ \bibinfo {pages} {L711}
  (\bibinfo {year} {1986})}\BibitemShut {NoStop}%
\bibitem [{\citenamefont {Orbach}\ and\ \citenamefont
  {Bleaney}(1961)}]{Orbach61}%
  \BibitemOpen
  \bibfield  {author} {\bibinfo {author} {\bibfnamefont {R.}~\bibnamefont
  {Orbach}}\ and\ \bibinfo {author} {\bibfnamefont {B.}~\bibnamefont
  {Bleaney}},\ }\href {\doibase 10.1098/rspa.1961.0212} {\bibfield  {journal}
  {\bibinfo  {journal} {Proceedings of the Royal Society of London. Series A.
  Mathematical and Physical Sciences}\ }\textbf {\bibinfo {volume} {264}},\
  \bibinfo {pages} {485} (\bibinfo {year} {1961})}\BibitemShut {NoStop}%
\bibitem [{\citenamefont {Zadrozny}\ \emph
  {et~al.}(2013{\natexlab{a}})\citenamefont {Zadrozny}, \citenamefont {Xiao},
  \citenamefont {Long}, \citenamefont {Atanasov}, \citenamefont {Neese},
  \citenamefont {Grandjean},\ and\ \citenamefont
  {Long}}]{zadrozny_mossbauer_2013}%
  \BibitemOpen
  \bibfield  {author} {\bibinfo {author} {\bibfnamefont {J.~M.}\ \bibnamefont
  {Zadrozny}}, \bibinfo {author} {\bibfnamefont {D.~J.}\ \bibnamefont {Xiao}},
  \bibinfo {author} {\bibfnamefont {J.~R.}\ \bibnamefont {Long}}, \bibinfo
  {author} {\bibfnamefont {M.}~\bibnamefont {Atanasov}}, \bibinfo {author}
  {\bibfnamefont {F.}~\bibnamefont {Neese}}, \bibinfo {author} {\bibfnamefont
  {F.}~\bibnamefont {Grandjean}}, \ and\ \bibinfo {author} {\bibfnamefont
  {G.~J.}\ \bibnamefont {Long}},\ }\href {\doibase 10.1021/ic402013n}
  {\bibfield  {journal} {\bibinfo  {journal} {Inorganic Chemistry}\ }\textbf
  {\bibinfo {volume} {52}},\ \bibinfo {pages} {13123} (\bibinfo {year}
  {2013}{\natexlab{a}})}\BibitemShut {NoStop}%
\bibitem [{\citenamefont {Lewis}\ and\ \citenamefont
  {Schwarzenbach}(1981)}]{Lewis:a19705}%
  \BibitemOpen
  \bibfield  {author} {\bibinfo {author} {\bibfnamefont {J.}~\bibnamefont
  {Lewis}}\ and\ \bibinfo {author} {\bibfnamefont {D.}~\bibnamefont
  {Schwarzenbach}},\ }\href {\doibase 10.1107/S0567739481001204} {\bibfield
  {journal} {\bibinfo  {journal} {Acta Crystallographica Section A}\ }\textbf
  {\bibinfo {volume} {37}},\ \bibinfo {pages} {507} (\bibinfo {year}
  {1981})}\BibitemShut {NoStop}%
\bibitem [{\citenamefont {Liu}\ \emph {et~al.}(2016)\citenamefont {Liu},
  \citenamefont {Chen}, \citenamefont {Liu}, \citenamefont {Vieru},
  \citenamefont {Ungur}, \citenamefont {Jia}, \citenamefont {Chibotaru},
  \citenamefont {Lan}, \citenamefont {Wernsdorfer}, \citenamefont {Gao},
  \citenamefont {Chen},\ and\ \citenamefont {Tong}}]{doi:10.1021/jacs.6b02638}%
  \BibitemOpen
  \bibfield  {author} {\bibinfo {author} {\bibfnamefont {J.}~\bibnamefont
  {Liu}}, \bibinfo {author} {\bibfnamefont {Y.-C.}\ \bibnamefont {Chen}},
  \bibinfo {author} {\bibfnamefont {J.-L.}\ \bibnamefont {Liu}}, \bibinfo
  {author} {\bibfnamefont {V.}~\bibnamefont {Vieru}}, \bibinfo {author}
  {\bibfnamefont {L.}~\bibnamefont {Ungur}}, \bibinfo {author} {\bibfnamefont
  {J.-H.}\ \bibnamefont {Jia}}, \bibinfo {author} {\bibfnamefont {L.~F.}\
  \bibnamefont {Chibotaru}}, \bibinfo {author} {\bibfnamefont {Y.}~\bibnamefont
  {Lan}}, \bibinfo {author} {\bibfnamefont {W.}~\bibnamefont {Wernsdorfer}},
  \bibinfo {author} {\bibfnamefont {S.}~\bibnamefont {Gao}}, \bibinfo {author}
  {\bibfnamefont {X.-M.}\ \bibnamefont {Chen}}, \ and\ \bibinfo {author}
  {\bibfnamefont {M.-L.}\ \bibnamefont {Tong}},\ }\href {\doibase
  10.1021/jacs.6b02638} {\bibfield  {journal} {\bibinfo  {journal} {Journal of
  the American Chemical Society}\ }\textbf {\bibinfo {volume} {138}},\ \bibinfo
  {pages} {5441} (\bibinfo {year} {2016})}\BibitemShut {NoStop}%
\bibitem [{\citenamefont {Ganivet}\ \emph {et~al.}(2013)\citenamefont
  {Ganivet}, \citenamefont {Ballesteros}, \citenamefont {de~la Torre},
  \citenamefont {Clemente-Juan}, \citenamefont {Coronado},\ and\ \citenamefont
  {Torres}}]{Ganivet2013}%
  \BibitemOpen
  \bibfield  {author} {\bibinfo {author} {\bibfnamefont {C.~R.}\ \bibnamefont
  {Ganivet}}, \bibinfo {author} {\bibfnamefont {B.}~\bibnamefont
  {Ballesteros}}, \bibinfo {author} {\bibfnamefont {G.}~\bibnamefont {de~la
  Torre}}, \bibinfo {author} {\bibfnamefont {J.~M.}\ \bibnamefont
  {Clemente-Juan}}, \bibinfo {author} {\bibfnamefont {E.}~\bibnamefont
  {Coronado}}, \ and\ \bibinfo {author} {\bibfnamefont {T.}~\bibnamefont
  {Torres}},\ }\href {\doibase 10.1002/chem.201202600} {\bibfield  {journal}
  {\bibinfo  {journal} {Chemistry – A European Journal}\ }\textbf {\bibinfo
  {volume} {19}},\ \bibinfo {pages} {1457} (\bibinfo {year}
  {2013})}\BibitemShut {NoStop}%
\bibitem [{\citenamefont {Zadrozny}\ \emph
  {et~al.}(2013{\natexlab{b}})\citenamefont {Zadrozny}, \citenamefont {Xiao},
  \citenamefont {Atanasov}, \citenamefont {Long}, \citenamefont {Grandjean},
  \citenamefont {Neese},\ and\ \citenamefont {Long}}]{zadrozny_magnetic_2013}%
  \BibitemOpen
  \bibfield  {author} {\bibinfo {author} {\bibfnamefont {J.~M.}\ \bibnamefont
  {Zadrozny}}, \bibinfo {author} {\bibfnamefont {D.~J.}\ \bibnamefont {Xiao}},
  \bibinfo {author} {\bibfnamefont {M.}~\bibnamefont {Atanasov}}, \bibinfo
  {author} {\bibfnamefont {G.~J.}\ \bibnamefont {Long}}, \bibinfo {author}
  {\bibfnamefont {F.}~\bibnamefont {Grandjean}}, \bibinfo {author}
  {\bibfnamefont {F.}~\bibnamefont {Neese}}, \ and\ \bibinfo {author}
  {\bibfnamefont {J.~R.}\ \bibnamefont {Long}},\ }\href {\doibase
  10.1038/nchem.1630} {\bibfield  {journal} {\bibinfo  {journal} {Nature
  Chemistry}\ }\textbf {\bibinfo {volume} {5}},\ \bibinfo {pages} {577}
  (\bibinfo {year} {2013}{\natexlab{b}})}\BibitemShut {NoStop}%
\bibitem [{\citenamefont {Kazin}\ \emph {et~al.}(2014)\citenamefont {Kazin},
  \citenamefont {Zykin}, \citenamefont {Zubavichus}, \citenamefont {Magdysyuk},
  \citenamefont {Dinnebier},\ and\ \citenamefont
  {Jansen}}]{doi:10.1002/chem.201303136}%
  \BibitemOpen
  \bibfield  {author} {\bibinfo {author} {\bibfnamefont {P.~E.}\ \bibnamefont
  {Kazin}}, \bibinfo {author} {\bibfnamefont {M.~A.}\ \bibnamefont {Zykin}},
  \bibinfo {author} {\bibfnamefont {Y.~V.}\ \bibnamefont {Zubavichus}},
  \bibinfo {author} {\bibfnamefont {O.~V.}\ \bibnamefont {Magdysyuk}}, \bibinfo
  {author} {\bibfnamefont {R.~E.}\ \bibnamefont {Dinnebier}}, \ and\ \bibinfo
  {author} {\bibfnamefont {M.}~\bibnamefont {Jansen}},\ }\href {\doibase
  10.1002/chem.201303136} {\bibfield  {journal} {\bibinfo  {journal} {Chemistry
  – A European Journal}\ }\textbf {\bibinfo {volume} {20}},\ \bibinfo {pages}
  {165} (\bibinfo {year} {2014})}\BibitemShut {NoStop}%
\bibitem [{\citenamefont {Giraud}\ \emph {et~al.}(2003)\citenamefont {Giraud},
  \citenamefont {Tkachuk},\ and\ \citenamefont
  {Barbara}}]{PhysRevLett.91.257204}%
  \BibitemOpen
  \bibfield  {author} {\bibinfo {author} {\bibfnamefont {R.}~\bibnamefont
  {Giraud}}, \bibinfo {author} {\bibfnamefont {A.~M.}\ \bibnamefont {Tkachuk}},
  \ and\ \bibinfo {author} {\bibfnamefont {B.}~\bibnamefont {Barbara}},\ }\href
  {\doibase 10.1103/PhysRevLett.91.257204} {\bibfield  {journal} {\bibinfo
  {journal} {Phys. Rev. Lett.}\ }\textbf {\bibinfo {volume} {91}},\ \bibinfo
  {pages} {257204} (\bibinfo {year} {2003})}\BibitemShut {NoStop}%
\bibitem [{\citenamefont {Giraud}\ \emph {et~al.}(2001)\citenamefont {Giraud},
  \citenamefont {Wernsdorfer}, \citenamefont {Tkachuk}, \citenamefont
  {Mailly},\ and\ \citenamefont {Barbara}}]{PhysRevLett.87.057203}%
  \BibitemOpen
  \bibfield  {author} {\bibinfo {author} {\bibfnamefont {R.}~\bibnamefont
  {Giraud}}, \bibinfo {author} {\bibfnamefont {W.}~\bibnamefont {Wernsdorfer}},
  \bibinfo {author} {\bibfnamefont {A.~M.}\ \bibnamefont {Tkachuk}}, \bibinfo
  {author} {\bibfnamefont {D.}~\bibnamefont {Mailly}}, \ and\ \bibinfo {author}
  {\bibfnamefont {B.}~\bibnamefont {Barbara}},\ }\href {\doibase
  10.1103/PhysRevLett.87.057203} {\bibfield  {journal} {\bibinfo  {journal}
  {Phys. Rev. Lett.}\ }\textbf {\bibinfo {volume} {87}},\ \bibinfo {pages}
  {057203} (\bibinfo {year} {2001})}\BibitemShut {NoStop}%
\bibitem [{\citenamefont {Banhart}\ \emph {et~al.}(1986)\citenamefont
  {Banhart}, \citenamefont {Ebert}, \citenamefont {Voitl\"ander},\ and\
  \citenamefont {Winter}}]{Banhart1986}%
  \BibitemOpen
  \bibfield  {author} {\bibinfo {author} {\bibfnamefont {J.}~\bibnamefont
  {Banhart}}, \bibinfo {author} {\bibfnamefont {H.}~\bibnamefont {Ebert}},
  \bibinfo {author} {\bibfnamefont {J.}~\bibnamefont {Voitl\"ander}}, \ and\
  \bibinfo {author} {\bibfnamefont {H.}~\bibnamefont {Winter}},\ }\href
  {\doibase 10.1016/0304-8853(86)90030-2} {\bibfield  {journal} {\bibinfo
  {journal} {J. Magn. Magn. Mater.}\ }\textbf {\bibinfo {volume} {61}},\
  \bibinfo {pages} {221} (\bibinfo {year} {1986})}\BibitemShut {NoStop}%
\bibitem [{\citenamefont {H\"ohn}\ \emph {et~al.}(2009)\citenamefont {H\"ohn},
  \citenamefont {Hoffmann}, \citenamefont {Hunger}, \citenamefont {Leoni},
  \citenamefont {Nitsche}, \citenamefont {Schnelle},\ and\ \citenamefont
  {Kniep}}]{Hohn2009}%
  \BibitemOpen
  \bibfield  {author} {\bibinfo {author} {\bibfnamefont {P.}~\bibnamefont
  {H\"ohn}}, \bibinfo {author} {\bibfnamefont {S.}~\bibnamefont {Hoffmann}},
  \bibinfo {author} {\bibfnamefont {J.}~\bibnamefont {Hunger}}, \bibinfo
  {author} {\bibfnamefont {S.}~\bibnamefont {Leoni}}, \bibinfo {author}
  {\bibfnamefont {F.}~\bibnamefont {Nitsche}}, \bibinfo {author} {\bibfnamefont
  {W.}~\bibnamefont {Schnelle}}, \ and\ \bibinfo {author} {\bibfnamefont
  {R.}~\bibnamefont {Kniep}},\ }\href {\doibase 10.1002/chem.200801857}
  {\bibfield  {journal} {\bibinfo  {journal} {Chem. Eur. J.}\ }\textbf
  {\bibinfo {volume} {15}},\ \bibinfo {pages} {3419} (\bibinfo {year}
  {2009})}\BibitemShut {NoStop}%
\bibitem [{\citenamefont {{Axtmann}}\ \emph {et~al.}(1968)\citenamefont
  {{Axtmann}}, \citenamefont {{Hazony}},\ and\ \citenamefont
  {{Hurley}}}]{1968CPL.....2..673A}%
  \BibitemOpen
  \bibfield  {author} {\bibinfo {author} {\bibfnamefont {R.~C.}\ \bibnamefont
  {{Axtmann}}}, \bibinfo {author} {\bibfnamefont {Y.}~\bibnamefont {{Hazony}}},
  \ and\ \bibinfo {author} {\bibfnamefont {J.~W.}\ \bibnamefont {{Hurley}},
  \bibfnamefont {Jr.}},\ }\href {\doibase 10.1016/0009-2614(63)80051-2}
  {\bibfield  {journal} {\bibinfo  {journal} {Chemical Physics Letters}\
  }\textbf {\bibinfo {volume} {2}},\ \bibinfo {pages} {673} (\bibinfo {year}
  {1968})}\BibitemShut {NoStop}%
\end{thebibliography}%
\end{document}